\documentclass{iopart}

\usepackage{iopams}
\usepackage{graphics,color,epsfig}
\renewcommand{\Im}{\mathop{\rm Im}}

\begin{document}
\title[Coherent Nonlinear Optics and Quantum Control in
Negative-Index Metamaterials]{Coherent Nonlinear Optics and
Quantum Control in Negative-Index Metamaterials}
\author{A K Popov$^1$, S A Myslivets$^2$ and V M Shalaev$^3$}
\address{$^1$ Department of Physics and Astronomy, University of
Wisconsin-Stevens Point, Stevens Point, WI 54481, USA \\
$^2$ Siberian Federal University and Institute of Physics of
Russian Academy of Sciences, 660036 Krasnoyarsk, Russian Federation\\
$^3$ Birk Nanotechnology Center and School of Electrical and
Computer Engineering, Purdue University, West Lafayette, IN 47907, USA}
\ead{apopov@uwsp.edu}
\begin{abstract}
The extraordinary properties of laser-induced transparency  of a negative-index slab and parametric amplification for a backward-wave signal are investigated. The effects of the idler absorption and phase mismatch on the amplification of the signal are studied, and the feasibility of ensuring robust transparency for a broad range of control field intensities and slab thicknesses is shown. A particular option consisting of the independent engineering of a strong four-wave mixing response and the negative refractive index is proposed and its specific features are investigated.  The feasibility of quantum control over the slab transparency in such a scheme is confirmed through numerical experiments. We thus show opportunities and conditions for the compensation of the strong absorption inherent to plasmonic negative-index metamaterials, and we further show achievable transparency through coherent energy transfer from the control optical field to the negative-index signal.
\end{abstract}
\vspace{2pc}
\maketitle

\section{Introduction} Negative-index (also known as  negative phase velocity or left-handed) metamaterials (NIMs) form a novel class of
electromagnetic media that promises revolutionary breakthroughs in
photonics \cite{Sh}. Significant progress has been achieved recently
in the design of bulk, multilayered, negative-index, plasmonic
structures \cite{Suk,Zh}. The majority of  NIMs realized to date
consist of metal-dielectric nanostructures that have highly
controllable magnetic and dielectric responses. The problem, however,
is that these structures have losses that are difficult to avoid,
especially in the visible range of frequencies.
Irrespective of their origin, losses constitute a major hurdle to the practical realization of the unique optical applications of these structures. Therefore, developing efficient loss-compensating techniques is of paramount importance. So far, the most common approaches to compensating losses in NIMs are related to the investigation of the possibility to embed amplifying centers in the host matrix \cite{Sh}. The amplification is supposed to be provided through a population inversion between the levels of the embedded centers. Herein, we investigate alternative options based on coherent, nonlinear optical (NLO) energy transfer from the control optical field(s) to the signal through optical parametric amplification (OPA).  Nonlinear optics in NIMs remains so far a less-developed branch of optics. On a fundamental level, the NLO response of nanostructured metamaterials is not completely understood or characterized and cannot be predicted effectively to date. Nevertheless, it is well established that local-field enhanced nonlinearities can be attributed to plasmonic nanostructures, and some rough estimates of their magnitude can be obtained. The feasibility of crafting NIMs with strong NLO responses in the optical wavelength range has been experimentally demonstrated in \cite{Kl}. Unlike natural, positive-index (PI) materials, the energy flow and the phase velocity are counter-directed in NIMs, which determines their extraordinary linear and NLO propagation properties. Unusual properties of nonlinear propagation processes in NIMs, such as second harmonic generation, three-wave mixing (TWM) and four-wave mixing (FWM)  OPA, which are in a drastic contrast with their counterparts in natural materials, were shown in \cite{Agr,Kiv,SHG,APB,Sc,OL,OLM,APB09,OL09,APL}. Striking changes in the properties of nonlinear pulse propagation and temporal solitons \cite{Laz}, spatial solitons in systems with bistability \cite{Tas,Kos,Boa}, gap solitons \cite{Agu}, and optical bistability in layered structures including NIMs \cite{Lit} were revealed.  A review of some of the corresponding theoretical approaches is given  in \cite{Gab}.

In the present paper, we investigate the effects of idler absorption and phase mismatch on parametric amplification of the backward waves and propose  a novel scheme of compensating losses based on the results of this investigation. The paper is organized as follows. In Section \ref{me}, the similarities between the solutions to the nonlinear propagation equations for the electric and magnetic, quadratic and cubic nonlinearities are shown for the case of uniform control fields. Here, the local linear and nonlinear parameters are assumed independent of the intensities of the control fields, and the conclusions are applicable to both TWM and FWM. The induced transparency  exhibits a resonance behavior as a function of the control field intensity and the NIM slab's thickness due to the backwardness of the light waves in NIMs. Usually, the resonances are narrow, especially if the parametric process is assisted by amplification of the idler due to Raman or population-inversion gain. The sample remains opaque anywhere beyond the resonance magnitudes of the control field and the resonance thickness of the sample. Counterintuitively, we show that transparency becomes achievable within a broad range of these parameters if the absorption for the idler exceeds that for the signal. Phase matching of the contrapropagating waves presents a technical challenge \cite{Har,Kh,Pas}. We also show that, in the above indicated case, the transparency of a NIM slab also becomes much more robust against the phase mismatch. Based on these outcomes, we propose and investigate in Section \ref{lta} the option of independent engineering of the negative index and FWM nonlinear response, which is different from the one proposed earlier \cite{OLM,APB09, OL09}. We consider the doped metamaterial, where the signal appears in the vicinity of the transition between the excited energy levels, whereas the idler couples with the ground state of the embedded quasiresonant centers that provide a resonantly enhanced, FWM response. Here, all local optical characteristics including nonlinear susceptibility exhibit a strong dependence on the intensity for the driving fields and on the frequency resonance offsets of the coupled waves. Hence, the output signal can be tailored through the means of quantum control. It is shown that the indicated cardinal changes in the coupling scheme bring about major changes in the properties of the laser-induced transparency of the doped NIM slabs.  The possibility of eliminating the negative role of the phase mismatch on the tailored transparency of the slab in this case is also shown. The results, supported by numerical simulations, prove the feasibility of tailored transparency, amplification and the creation of a microscopic, mirrorless, backward-wave, optical parametric oscillator that generates contradirected beams of entangled right- and left-handed photons.

\section{{Backward waves, parametric interaction in a  NIM and solutions to the nonlinear propagation equations }}\label{me}
\subsection{{Poynting and wave-vectors in a lossless
NIM}}\label{pv}
We consider a traveling electromagnetic wave,
\begin{eqnarray}
\mathbf{E}(\mathbf{r},t)&=&(1/2)\mathbf{E}_0\exp[i(\mathbf{
k\cdot r}-\omega t)]+ c.c.,  \label{EM} \\
\mathbf{H}(\mathbf{r},t)&=&(1/2)\mathbf{H}_0\exp[i(\mathbf{
k\cdot r}-\omega t)]+ c.c.  \label{HM}
\end{eqnarray}
From the  equations
\begin{equation}\nabla \times \mathbf{E} = -\frac{1}{c}\frac{\partial
\mathbf{B}}{\partial t}, \mathbf{B} =\mu \mathbf{H}, \nabla \times \mathbf{H}=
\frac{1}{c}\frac{\partial \mathbf{D}}{\partial t}, \mathbf{D} =\epsilon \mathbf{E} \label{ME1}
\end{equation}
one finds that
\begin{equation}
\mathbf{k}\times\mathbf{E} = \frac{\omega}{c} \mu\mathbf{H},
\mathbf{k}
\times\mathbf{H} =-\frac{\omega}{c} \epsilon\mathbf{E},
\sqrt{\epsilon}{E}=-\sqrt{\mu}{H}.
\label{eh}
\end{equation}
Equations (\ref{eh}) show that
the vector triplet $\mathbf{E}$, $\mathbf{H}$, $\mathbf{k}$
forms a right-handed system for an ordinary medium with
$\epsilon_i>0$ and $\mu_i>0$.  Simultaneously negative $\epsilon_i$ and
$\mu_i$ result in a left-handed triplet and negative refractive index
\begin{equation}
n= - \sqrt{\mu\epsilon},\quad
{k}^{2}=n^{2}(\omega/{c})^{2}. \label{n}
\end{equation}
We assume here that all indices
of $\epsilon$, $\mu$
and $n$ are real numbers. The direction of the  wave-vector $\mathbf{k}$ with respect to the energy flow (Poynting vector) depends on the signs of $\epsilon $
and $\mu $:
\begin{equation}
\mathbf{S}(\mathbf{r},t)
=\frac{c}{4\pi}[\mathbf{E}\times\mathbf{H}] =\frac{c^{2}\mathbf{k}}{4\pi\omega\epsilon}H^{2}
=\frac{c^{2}\mathbf{k}}{4\pi\omega\mu}E^{2}.  \label{s}
\end{equation}
At $\epsilon_i<0$ and $\mu_i<0$, $\mathbf{S}$ and $\mathbf{k}$ become contradirected,
which is in contrast with the electrodynamics of ordinary
media and opens opportunities for many revolutionary
breakthroughs.
\subsection{{Coupling geometry and coherent energy transfer
from the ordinary control fields to the  backward signal
in a  NIM}}\label{opa}
\begin{figure}[!h]
\begin{center}
\includegraphics[width=.32\columnwidth]{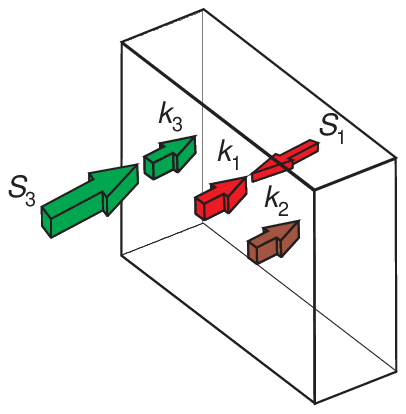}
\includegraphics[width=.32\columnwidth]{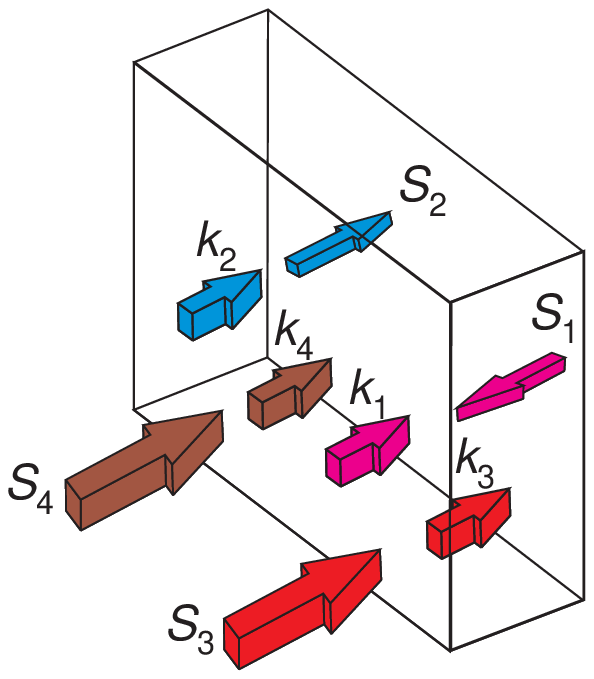}
\includegraphics[width=.32\columnwidth]{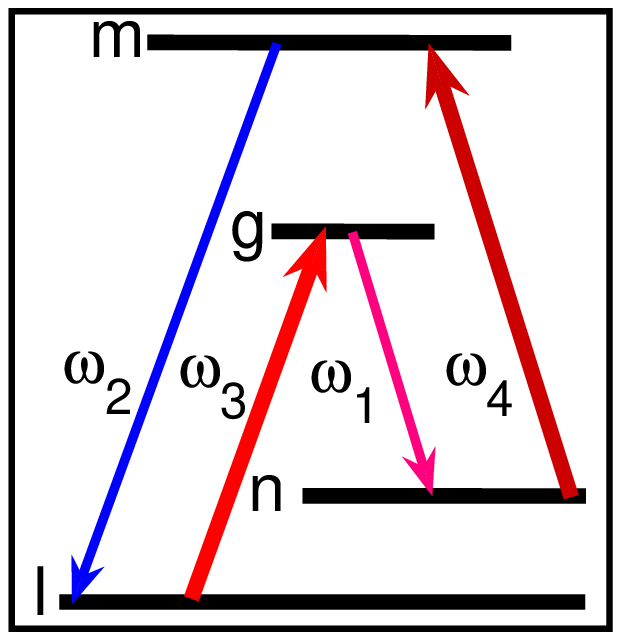}\\
(a)\hspace{30mm} (b)\hspace{30mm} (c)
\end{center}
\caption{Coupling geometry for three-wave mixing (a),  four-wave mixing (b) of the ordinary  and one backward electromagnetic waves  and scheme of quantum controlled four-wave mixing in the embedded resonant nonlinear-optical centers (c). Here, $\mathbf{S}_1$,
$\mathbf{k}_1$ and  $\omega_1$ are energy flow, wave-vector and frequency of the negative-index signal ($n_1<1$);  $\mathbf{S}_{3,4}$,
$\mathbf{k}_{3,4}$ and $\omega_{3,4}$ - of the positive-index control fields;
$\mathbf{k}_{2}$ and $\omega_{2}$ stand for the positive-index idler. }
\label{f1}
\end{figure}
The basic idea of compensating losses by coherent energy
transfer from the control field(s) to the signal through parametric interaction is illustrated in
Fig. \ref{f1}.
We assume the wave at $\omega_{1}$ with the wave-vector
$\mathbf{k}_1$
directed along the $z$-axis is a negative-index (NI) ($n_{1}<0$)
signal. Therefore, it is a backward wave because its energy flow $\mathbf{S}_{1}
=(c/4\pi)[\mathbf{E_1}\times \mathbf{H_1}]$ appears directed against the
$z$-axis. In the TWM case, shown in Fig. \ref{f1}(a), the medium is illuminated by a higher-frequency, ordinary  PI wave at $\omega_{3}$
traveling along the $z$-axis ($n_{3}>0$). In the FWM case, Fig. \ref{f1}(b),  the slab is illuminated by two PI control (pump)  waves at $\omega_{3}$ and  $\omega_{4}$. In both cases, all wave-vectors are co-directed along the the $z$-axis. Due to the parametric interaction, the control and signal fields generate a difference-frequency idler at
$\omega_{2}=\omega_{3}-\omega_{1}$ (TWM) or at $\omega_{2}=\omega_{4}+\omega_{3}-\omega_{1}$ (FWM), which is
also assumed to be a PI wave ($n_{2}>0$). The idler contributes
back into the wave at $\omega_{1}$ through the same type of the parametric interaction and thus enables OPA at $\omega_{1}$ by converting the energy of the control fields into the signal. Thus, all of the
coupled waves have their wave-vectors {co-directed} along $z$, whereas
the energy flow of the signal wave, $\mathbf{S}_{1}$, is
{counter-directed} to the energy flows of all the other waves, which are codirected with their wave-vectors. Such coupling schemes are in contrast both with the conventional phase-matching scheme for OPA in ordinary materials, where all energy-flows and phase velocities are co-directed, as well as with TWM backward-wave mirrorless OPO \cite{Har,Kh,Pas,Vol}, where {both} the energy flow {and} wave-vector of one of the waves are opposite to all others.

\subsection{{Equations for coupled contrapropagating backward and ordinary waves }}\label{np}
First, we shall show that magnetic and electric TWM and FWM processes can be treated identically.
We consider two alternative types of nonlinearities --
{electric}, $\mathbf{D} = \epsilon \mathbf{E} + 4\pi
\mathbf{P}^{NL}$, $\mathbf{B} = \mu \mathbf{H}$; and
{magnetic}, $\mathbf{B} = \mu \mathbf{H} + 4\pi
\mathbf{M}^{NL}$, $\mathbf{D} =\epsilon \mathbf{E}$. Nonlinear
polarization and magnetization are sought in the form
\begin{eqnarray}
\mathbf{P}^{NL}(\mathbf{r},t)=(1/2)\mathbf{P_0}^{NL}(\mathbf{r})\exp[i(\mathbf{
\widetilde{k}\cdot r}-\omega t)]+ c.c., \quad \label{P} \\
\mathbf{M}^{NL}(\mathbf{r},t)=(1/2)\mathbf{M_0}^{NL}(\mathbf{r})\exp[i(\mathbf{
\widetilde{k}\cdot r}-\omega t)]+ c.c.\quad  \label{M}
\end{eqnarray}
Accounting for Eqs. (\ref{eh}), one can
derive
\begin{eqnarray}
\triangledown \times\nabla \times \mathbf{E} =
-(\mu/c^2)\partial^2 \mathbf{D}/\partial t^2,\nonumber\\
-\triangle \mathbf{E} = \mu(\omega^2/c^2)[\epsilon \mathbf{E}
+ 4\pi \mathbf{P}^{NL}],
\label{ME21}\\
\nabla \times\nabla \times \mathbf{H} =
-(\epsilon/c^2)\partial^2 \mathbf{B}/\partial t^2,\nonumber\\
-\triangle \mathbf{H} = \epsilon(\omega^2/c^2)[ \mu\mathbf{H}
+ 4\pi \mathbf{M}^{NL}].
\label{ME22}
\end{eqnarray}
For the  medium with the electric nonlinearity, the equation for the slowly varying
amplitude $\mathbf{E}_0$ of the wave with the wave-vector along
the $z$-axis takes the form:
\begin{equation}
d E_0/d
z=i\mu(2\pi\omega^2/kc^2)P_0^{NL}\exp[i(\widetilde{k}-k)z].\label{E0}
\end{equation}
For the magnetic nonlinearity, the  equation is
\begin{equation}
d H_0/d
z=i\epsilon(2\pi\omega^2/kc^2)M_0^{NL}\exp[i(\widetilde{k}-k)z].\label{H0}
\end{equation}
The equations are symmetric and can be converted from one to the other by
replacing $\mu \longleftrightarrow \epsilon$.

For the electric quadratic nonlinearity,
\begin{eqnarray}
P_{1}^{NL} &=&\chi_{e1}^{(2)}E_{3}E_{2}^{\ast }\exp
\{i[(k_{3}-k_{2})z-\omega_{1}t]\},  \label{p1} \\
P_{2}^{NL} &=&\chi_{e2}^{(2)}E_{3}E_{1}^{\ast }\exp
\{i[(k_{3}-k_{1})z-\omega_{2}t]\},  \label{p2}
\end{eqnarray}
where  $\omega_{2}=\omega_{3}-\omega_{1}$ and
$k_{j}=|n_{j}|\omega _{j}/c>0$. Then the equations for the
slowly-varying amplitudes of the signal and idler in the lossy medium can be given in the form
\begin{eqnarray}
{dE_{1}}/{dz} &=&i\sigma_{e1}E_{2}^{\ast }\exp [i\Delta kz]+({%
\alpha _{1}}/{2})E_{1},  \label{e1} \\
{dE_{2}}/{dz} &=&i\sigma_{e2}E_{1}^{\ast }\exp [i\Delta kz]-({%
\alpha_{2}}/{2})E_{2}.  \label{e2}
\end{eqnarray}
Here,
$\sigma_{ej}=(k_{j}/\epsilon_j)2\pi\chi_{ej}^{(2)}E_3$,
$\Delta k=k_{3}-k_{2}-k_{1}$, and $\alpha_{j}$ are the
absorption indices. The amplitude of the control (pump) wave $E_{3}$ is assumed constant.

For the magnetic type of quadratic nonlinearity,
\begin{eqnarray}
M_{1}^{NL} &=&\chi_{m1}^{(2)}H_{3}H_{2}^{\ast }\exp
\{i[(k_{3}-k_{2})z-\omega_{1}t]\},  \label{m1} \\
M_{2}^{NL} &=&\chi_{m2}^{(2)}H_{3}H_{1}^{\ast }\exp
\{i[(k_{3}-k_{1})z-\omega_{2}t]\},  \label{m2}
\end{eqnarray}
the equations for the slowly-varying amplitudes are:
\begin{eqnarray}
{dH_{1}}/{dz} &=&i\sigma_{m1}H_{2}^{\ast }\exp [i\Delta kz]+{%
(\alpha_{1}}/{2})H_{1},  \label{h11} \\
{dH_{2}}/{dz} &=&i\sigma_{m2}H_{1}^{\ast }\exp [i\Delta kz]-{%
(\alpha_{2}}/{2})H_{2}.  \label{h21}
\end{eqnarray}
Here, $\sigma_{mj}=
(k_{j}/\mu_j)2\pi\chi_{mj}^{(2)}H_3$, $H_{3}= const$, and the other notations remain the same.

For the electric-type FWM, the equations for the
slowly-varying amplitudes are similar:
\begin{eqnarray}
{dE_{1}}/{dz} &=&i\gamma_1 E_{2}^{\ast }\exp [i\Delta
kz]+({\alpha_{1}%
}/{2} )E_{1},  \label{1} \\
{dE_{2}}/{dz} &=&i\gamma_2 E_{1}^{\ast }\exp [i\Delta
kz]-({\alpha_{2}}/{2 })E_{2}.  \label{2}
\end{eqnarray}
Here, $\gamma_{j}= (k_{j}/\epsilon_{j}){2\pi}
\chi_{j}^{(3)}E_{3}E_{4}$ and $\Delta k=k_{3}+k_{4}-k_{1}-k_{2}$.

We note the following {three fundamental differences} in equations (\ref{e1})-(\ref{2}) as compared with their counterpart in ordinary, PI materials. First, the signs of $\sigma_{1}$ and $\gamma_1$ are opposite to those of $\sigma_{2}$ and $\gamma_2$ because $\epsilon_{1}<0$ and $\mu_{1}<0$. Second, the opposite sign appears with $\alpha_{1}$ because the energy flow $\mathbf{S_{1}}$ is  against the $z$-axis. Third, the boundary conditions for the signal  are defined at the opposite side of the sample as compared to the idler  because the energy flows $\mathbf{S_{1}}$ and $\mathbf{S_{2}}$ are counter-directed.

We introduce effective  amplitudes, $a_{e,m,j}$, and
nonlinear coupling parameters, $g_{e,m,j}$, which for the
{electric} and magnetic types of quadratic nonlinearity are defined as
\begin{eqnarray}
a_{ej}=\sqrt{|\epsilon_j/k_j|}E_j,
g_{ej}=\sqrt{|k_1k_2/\epsilon_1\epsilon_2|}  2\pi\chi^{(2)}_{ej}E_3,\quad
\label{ec}\\
a_{mj}=\sqrt{|\mu_j/k_j|}H_j,
g_{mj}=\sqrt{|k_1k_2/\mu_1\mu_2|}  2\pi
\chi^{(2)}_{mj}H_3,\quad \label{mc}
\end{eqnarray}
and for  FWM as
\begin{equation}
a_{j}=\sqrt{|\epsilon_j/k_j|}E_j, g_j=\sqrt{|k_{1}k_{2}/\epsilon_{1}\epsilon_{2}|}
{2\pi}\chi_{j}^{(3)}E_{3}E_{4}.
\label{ga2}
\end{equation}
The quantities  $|a_j|^2$ are proportional to the photon numbers in the energy fluxes.
Equations for amplitudes $a_j$ are identical for all of the
types of nonlinearities studied here:
\begin{eqnarray}
{da_{1}}/{dz}&=&-g_{1}a^{\ast }_{2}\exp(i\Delta k
z)+({\alpha_1}/{2})a_{1},\label{a1}\\
{da_{2}}/{dz}&=&g_{2}a^{\ast }_{1}\exp(i\Delta k
z)-({\alpha_2}/{2})a_{2}.\label{a2}
\end{eqnarray}
\subsection{{Manley-Rowe relations and solutions to the equations  for coupled counter-propagating waves}}
\label{mrr}
At $\alpha_{1,2}=0$, $g_1=g_2$, e.g., for off-resonant coupling, one finds with the aid of equations
(\ref{s}) and (\ref{a1}), (\ref{a2}):
\begin{equation}
\frac{d}{dz}\left[ \frac{S_{1z}}{\hbar
{\omega_{1}}}-\frac{S_{2z}}{%
\hbar {\omega_{2}}}\right] =0,\quad
\frac{d}{dz}\left[|a_1|^2+|a_2|^2\right] =0.  \label{MR}
\end{equation}
These equations represent the Manley-Rowe relations \cite{MR}, which describe the creation of pairs of {entangled counter-propagating photons} $\hbar{\omega_{1}}$ and $\hbar{\omega_{2}}$. The second equation predicts that the {sum} of the terms proportional to the squared amplitudes of the signal and idler remains constant through the sample, which is due to the opposite signs of $S_{1z}$ and $S_{2z}$ and is in contrast with the requirement that the {difference} of such terms is constant in the analogous case in ordinary nonlinear-optical materials.

Taking into account the boundary conditions $a_{1}(z=L)=a_{1L}$,
and
$a_{2}(z=0)=a_{20}$ ($L$ is the slab thickness), the
solutions
to equations (\ref{a1}), (\ref{a2}) can be written as
\begin{eqnarray}
a_{1}(z)&=&A_{1}\exp [(\beta_{1}+i\frac{\Delta k}{2})z]+\nonumber\\
&+&A_{2}\exp [(\beta_{2}+i\frac{\Delta k}{2})z],  \label{a1z1}
\\
a_{2}^{\ast }(z)&=&\kappa_{1}A_{1}\exp
[(\beta_{1}-i\frac{\Delta
k}{2})z]+\nonumber\\
&+&\kappa_{2}A_{2}\exp [(\beta_{2}-i\frac{\Delta k}{2})z],
\label{a2z1}
\end{eqnarray}
where
\begin{eqnarray}
\beta_{1,2}={(\alpha_{1}-\alpha_{2})}/{4}\pm
iR,\,\kappa_{1,2}=[\pm {R}+is]/g, \quad\label{bet1} \\
R=\sqrt{g^{2}-s^{2}}, g^2=g_2^{\ast}g_1,  s=({\alpha_{1}+\alpha_{2}})/{4}-i{\Delta
k}/{2},\quad\label{r1} \\
A_{1}=\{a_{1L}\kappa_{2}-a_{20}^{\ast}\exp
[(\beta_{2}+i\frac{\Delta k}{2})L]\}/D,\quad \label{A11} \\
A_{2}=-\{a_{1L}\kappa_{1}-a_{20}^{\ast}\exp
[(\beta_{1}+i\frac{\Delta k}{2})L]\}/D,\quad  \label{A21} \\
D=\kappa_{2}\exp [(\beta_{1}+i\frac{\Delta
k}{2})L]-\kappa_{1}\exp
[(\beta_{2}+i\frac{\Delta k}{2})L].\quad \label{D1}
\end{eqnarray}
At $\Delta k=0$ and $\Im g=0$,
$(\alpha_{1}+\alpha_{2})L\ll \pi$ (off-resonance), equations (\ref{a1z1}) and (\ref{a2z1}) reduce to
\begin{eqnarray}
a_{1}^{\ast }(z) &\approx &\frac{a_{1L}^{\ast }}{\cos (gL)}\cos
({gz})+\frac{
ia_{20}}{\cos (gL)}\sin [{g(z-L)}],\qquad \label{lla11} \\
a_{2}(z) &\approx &\frac{ia_{1L}^{\ast }}{\cos (gL)}\sin
({gz})+\frac{a_{20} }{\cos (gL)}\cos [{g(z-L)}].\qquad
\label{lgen21}
\end{eqnarray}
The output amplitudes are then given by
\begin{eqnarray}
a_{10}^{\ast } &=&[{a_{1L}^{\ast }}/{\cos (gL)}]-ia_{20}\tan
({gL}),
\label{a101} \\
a_{2L} &=&ia_{1L}^{\ast }\tan ({gL})+[{a_{20}}/{\cos
({gL})}].
\label{a2l1}
\end{eqnarray}
At $a_{20}=0$,  the equations for the {energy distribution for the
backward wave},  $T_1(z)=\left\vert
{a_{1}(z)}/{a_{1L}}\right\vert ^{2}$, and for the  PI idler,
$\eta_{2}(z)=\left\vert {a_{2}(z)}/{a_{1L}^{\ast }}\right\vert
^{2}$, across the slab take the form
\begin{eqnarray}\label{eta4}
T_{1}(z)&=&\left|[{\kappa_2
    \exp{(\beta_1z)}-\kappa_1\exp{(\beta_2z)}}]/D\right|^2,\qquad\label{tz}\\
\eta_{2}(z)&=&\left|[\exp{(\beta_1z)}-\exp{(\beta_2z)}]/{D}\right|^2.\qquad\label{ez}
\end{eqnarray}
Then the transmission factor for the backward-wave signal at $z=0$, $T_{10}$, and the output idler at $z=L$, $\eta_{2L}$,  are given by
\begin{eqnarray}
T_{10}&=&\left\vert\frac
{a_{1}(0)}{a_{1L}}\right\vert ^{2}=\left|\frac{\exp \left\{-\left[ \left(
\alpha_{1}/2\right)-s%
\right] L\right\}}{\cos RL+\left(s/R\right) \sin RL}\right|^2,
\label{T}\\
\eta_{2L}&=&\left\vert\frac
{a_{2}(L)}{a_{1L}}\right\vert ^{2}=\left|\frac{(g/R)\sin RL}{\cos RL+\left(s/R\right) \sin RL}\right|^2.
\label{et}
\end{eqnarray}
At $a_{1L}=0$, $a_{2}(z=0)=a_{20}$, the slab serves as an NLO mirror with a reflectivity (output conversion efficiency) at $\omega_1$ given by an equation identical to Eq.(\ref{et}):
\begin{equation}
\eta_{10}=\left\vert\frac
{a_{1}(0)}{a_{20}^{\ast
}}\right\vert ^{2}=\left|\frac{(g/R)\sin RL}{\cos RL+\left(s/R\right) \sin RL}\right|^2. \label{gen11}
\end{equation}
\section{Laser-induced transparency, amplification and
generation of the backward wave}\label{lta}
The fundamental difference between the spatial distribution of
the signal in ordinary and NI slabs is explicitly seen at
$\alpha_j=\Delta k=0$. There, equation (\ref{T}) reduces to
\begin{equation}
T_{10}=1/[\cos(gL)]^2.  \label{T1}
\end{equation}
Equations (\ref{lla11})-(\ref{T1}) show that the output signal and
idler experience  a sequence of {geometrical} resonances at $g L
\rightarrow (2j+1)\pi/2$, (j=0, 1, 2, ...), as functions of the slab
thickness $L$ and of the intensity of the control field (factor $g$).
Such behavior is in drastic contrast with that in an ordinary medium,
where the signal would grow exponentially as $ T_1\propto \exp(2gL)$.
The resonances indicate that strong absorption  of the left-handed
wave and of the idler can be turned into transparency, amplification
and even into {cavity-free self-oscillation} when the
denominator tends to zero. The conversion factors $\eta_{10}$ and $\eta_{20}$ experience a similar resonance increase. Self-oscillations would provide for the generation of {entangled counter-propagating
left-handed, $\hbar\omega_1$, and right-handed, $\hbar\omega_2$,
photons without a cavity}.  A similar behavior is characteristic for
distributed-feedback lasers and is equivalent to a great extension of
the NLO coupling length. It is known that even weak amplification per
unit length may lead to lasing provided that the corresponding
frequency coincides with high-quality cavity or feedback
resonances. Below, we present investigation of the properties of such resonances.
\subsection{Effect of the idler absorption and phase mismatch on the laser-induced transparency resonances}\label{a}
\begin{figure}[h]
\begin{center}
\includegraphics[width=.445\columnwidth]{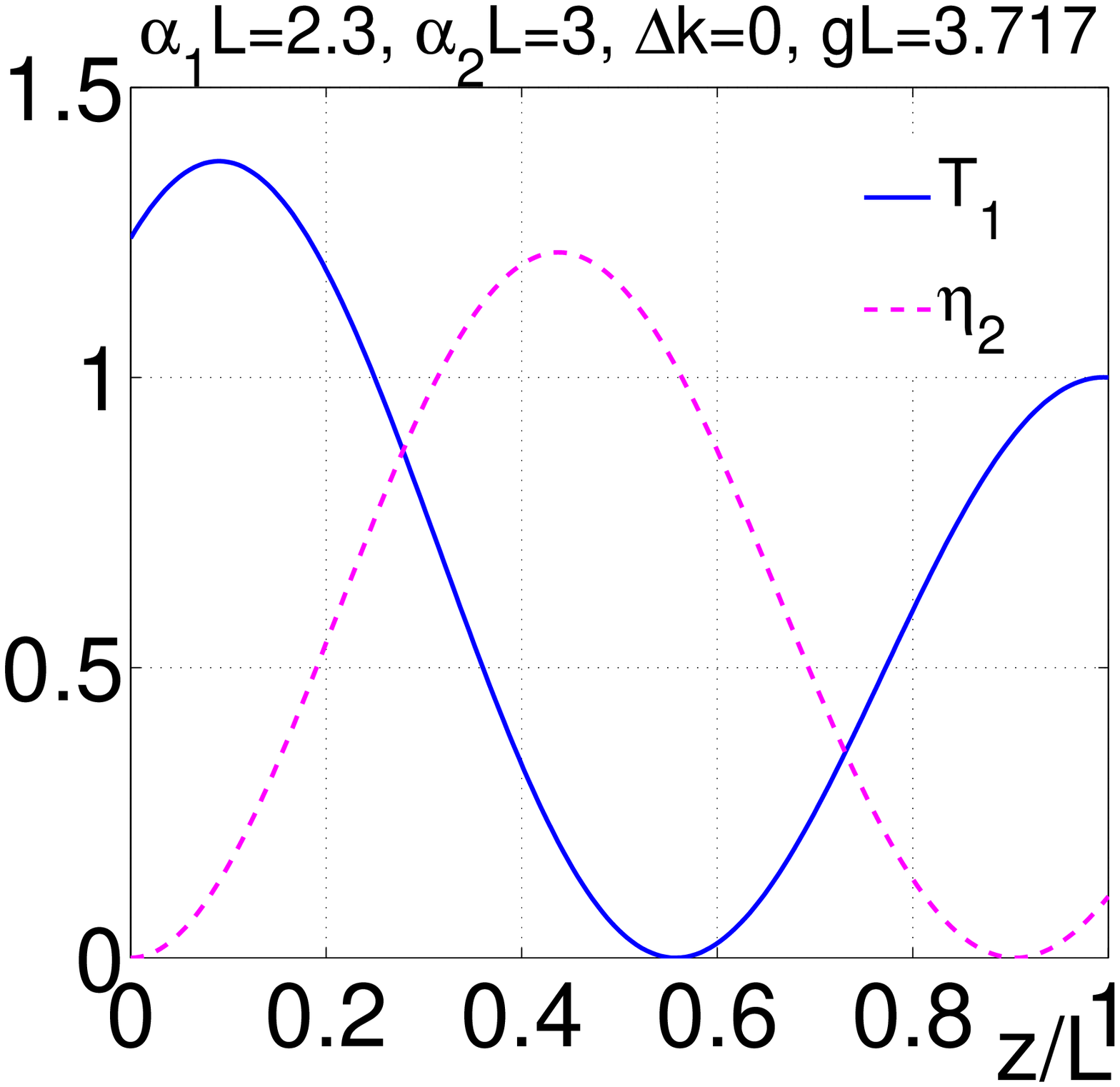}
\includegraphics[width=.445\columnwidth]{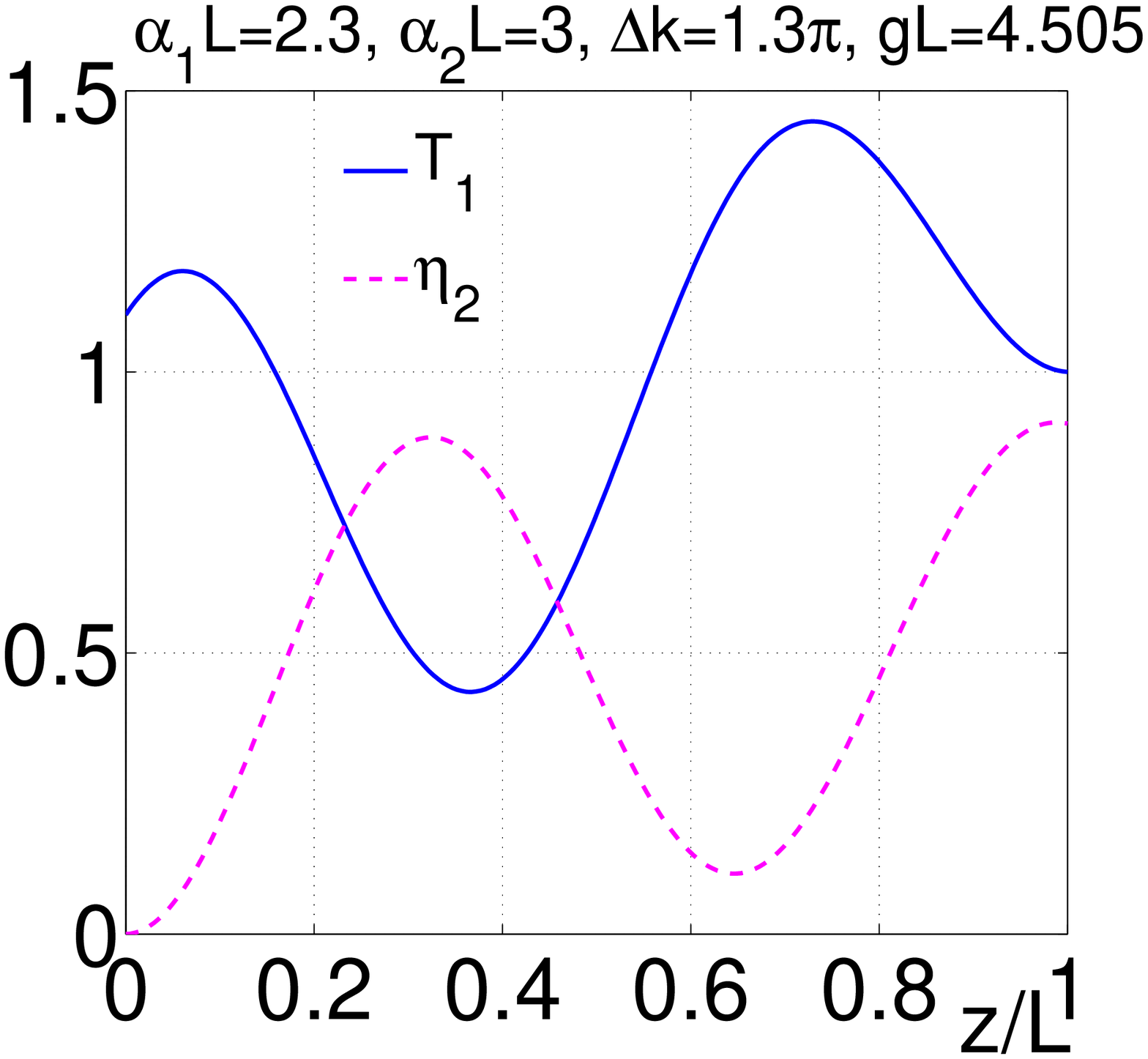}\\
(a)\hspace{30mm} (b)
\end{center}
\caption{ Tailored distribution of the signal, $T_1(z)=|a_1(z)/a_1(L)|^{2}$,
and of the idler, $\protect\eta_2(z)=|a_2(z)/a_1(L)|^2$, along the slab
at $\protect\alpha_{1}L=2.3$, $\protect\alpha_{2}L= 3$ in the first
transmission minimum.
(a): $\Delta k=0$, $gL=3,717$. (b): $\Delta k=1.3 \pi$, $gL=4,505$. }
\label{f2}
\end{figure}

\begin{figure}[h]
\begin{center}
\includegraphics[width=.445\columnwidth]{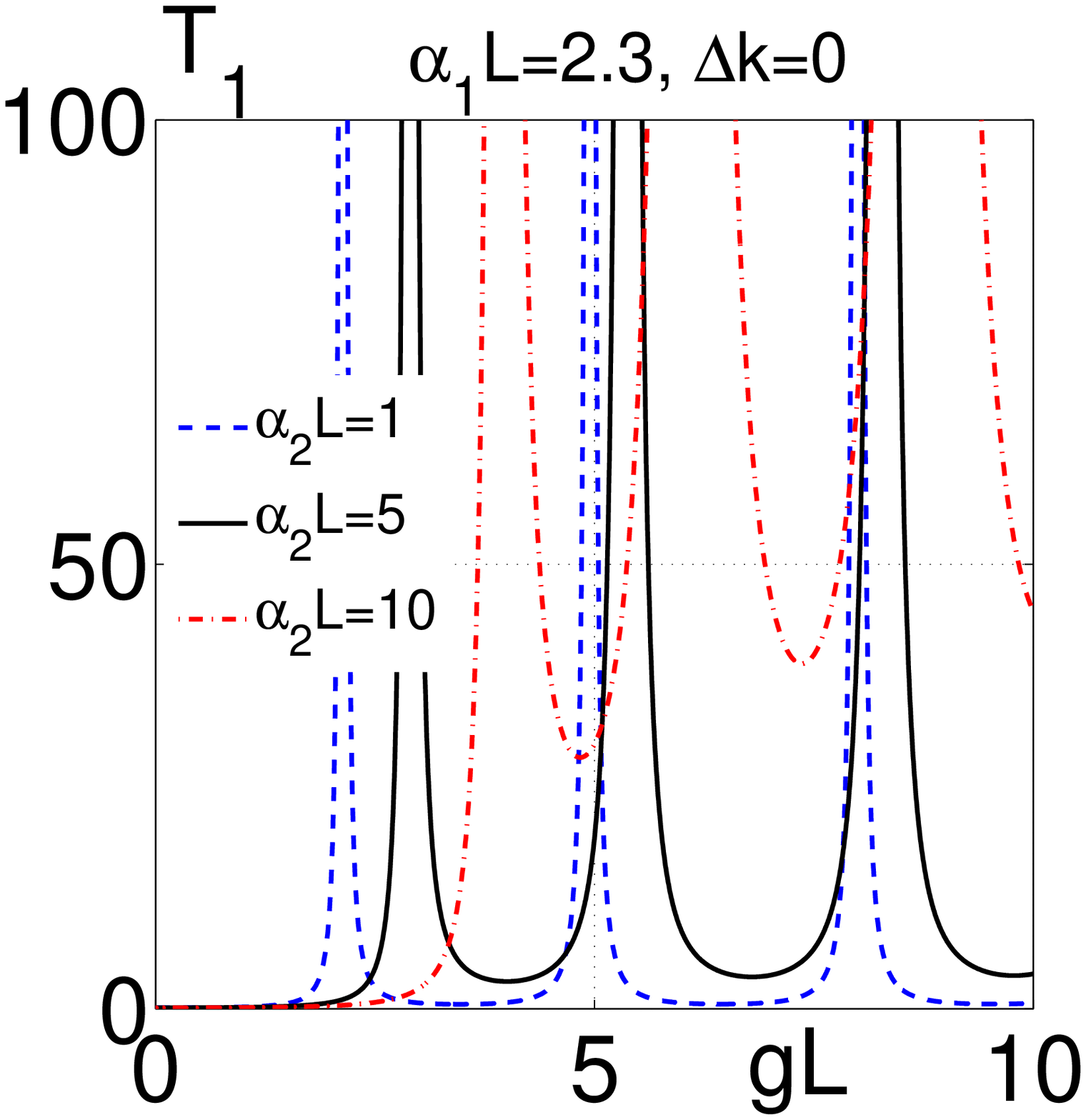}
\includegraphics[width=.445\columnwidth]{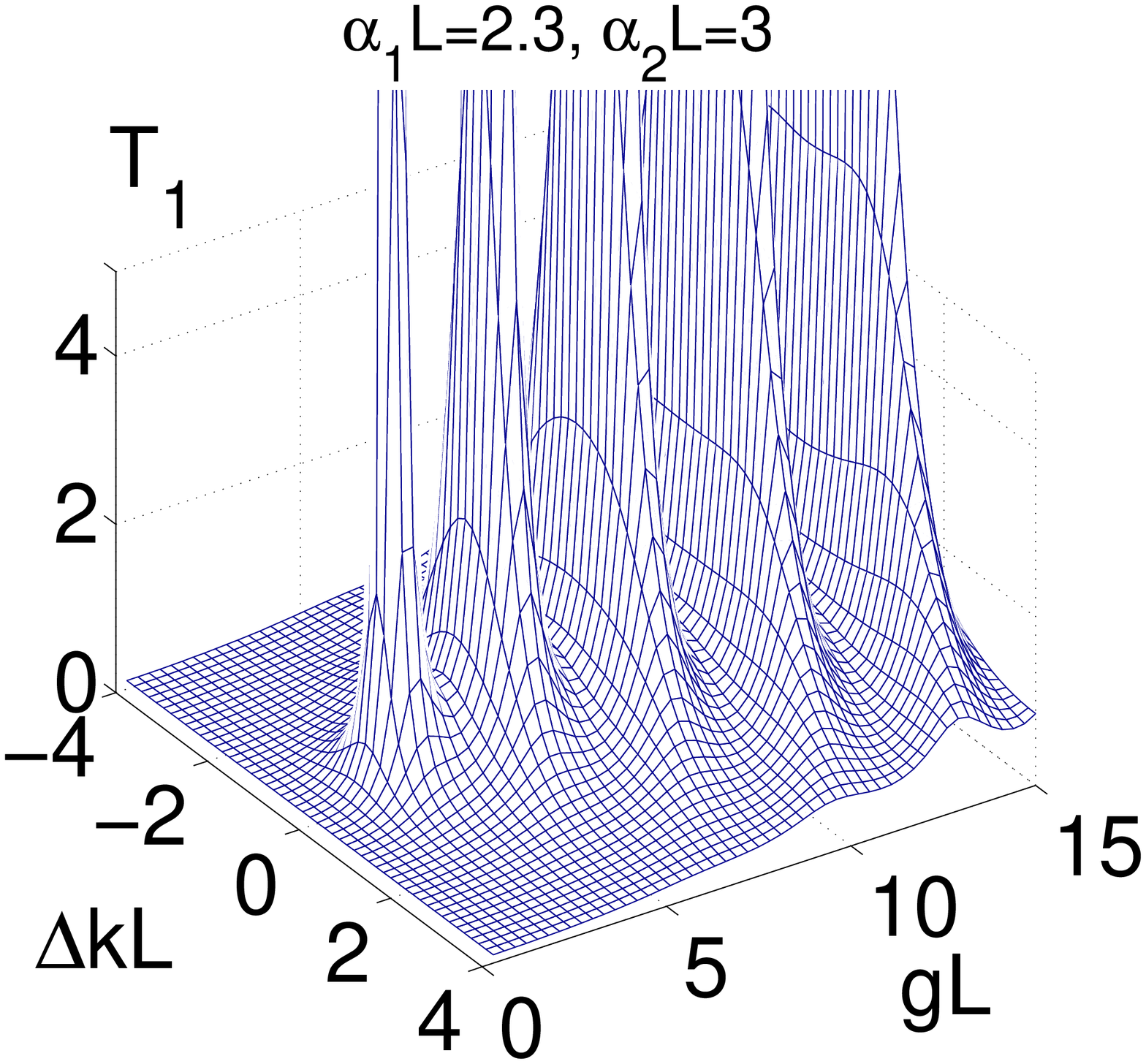}\\
(a)\hspace{30mm} (b) \\
\includegraphics[width=.445\columnwidth]{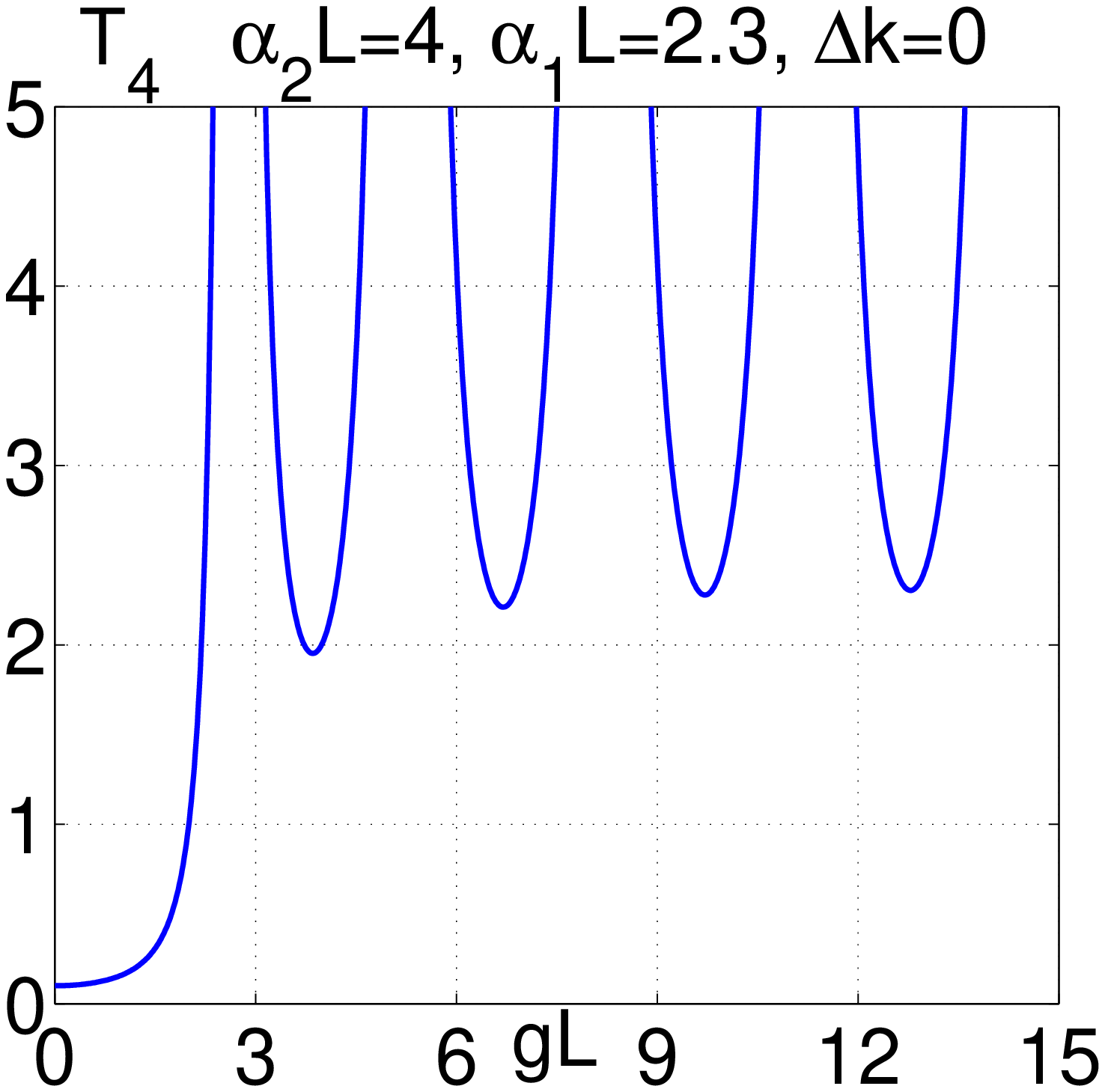}
\includegraphics[width=.445\columnwidth]{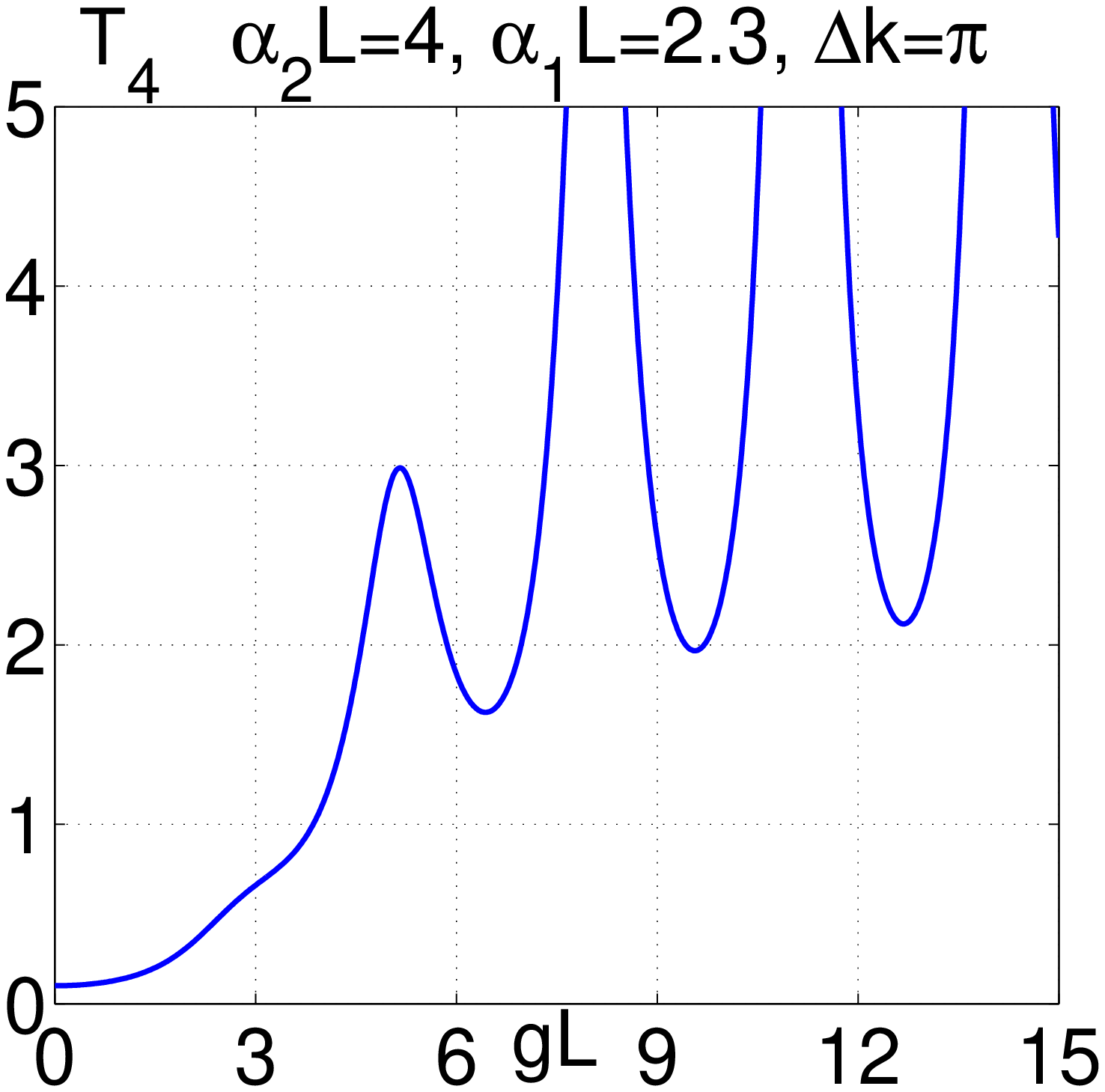}\\
(c)\hspace{30mm} (d)
\end{center}
\caption{Transmission of the negative-index slab, $T_{1}(z)$, vs.
parametric gain, $gL$, at $\alpha_1L=2.3$ and different values of
$\alpha_2L$ and $\Delta kL$. [(a), (c)] $\Delta k=0$; [(d)] $\Delta k=\pi$; [(c), (d)]  $\alpha_2L=4$.}
\label{f3}
\end{figure}

In order to demonstrate the major effects of the idler absorption and phase mismatch on the laser-induced transparency resonances, we consider the model where the dependence of the local optical and NLO parameters on the intensity of the control field can be neglected and the parameter $g$ is real. Such a model is relevant to, e.g., off-resonant quadratic and cubic nonlinearities attributed to the structural elements of metal-dielectric nanocomposites \cite{Kl}.  The results will be used in the next subsection for optimization of the transparency achievable through embedded resonant FWM centers with power-dependent optical parameters. A crucial role of the outlined geometrical resonances is illustrated in Figs.~\ref{f2} and \ref{f3}. Besides the factor $g$, the local NLO energy conversion rate for each of the waves is proportional to the amplitude of another coupled wave and depends on the  phase mismatch $\Delta k$. Hence, the fact that the waves decay in opposite directions causes a specific, strong dependence of the entire propagation process and, consequently, of the transmission properties of the slab on the ratio of the decay rates \cite{APL}.  A typical NIM slab absorbs about 90\% of light at the frequencies which are in the NI frequency range. Such absorption corresponds to $\alpha_1L\approx 2.3$. Since the idler grows toward the back facet of the slab  and the signal experiences absorption in the opposite direction, the maximum of the signal for the given parameters is located closer to the back facet of the slab. A change in the slab thickness or in the intensity of the control fields leads to significant changes in the distributions of the signal and idler along the slab and in their output values (Fig.~\ref{f2}).
As outlined above, the transparency exhibits an extraordinary
resonance behavior as a function of the intensity of the control field
and the NIM slab thickness, which occurs due to the backwardness of
the light waves in NIMs and is illustrated in Fig. ~\ref{f3}.
Basically, such resonances are narrow, like those depicted in the plot corresponding to $\alpha_2L=1$ in Fig. ~\ref{f3}(a), and the sample remains opaque anywhere beyond the resonance field and the parameters of the sample. If nonlinear susceptibility varies within the negative-index frequency domain, this translates into relatively narrow-band filtering. The slab would become transparent within the broad range of the slab thickness and the control field intensity if the transmission in all of the minimums is about or more than 1. Figure~\ref{f3} shows the feasibility of achieving {robust transparency and amplification in a NIM slab at the signal frequency through a wide range of the control field intensities and slab thicknesses} by the appropriate adjustment of the absorption indices $\alpha_{2}\geq\alpha_{1}$. Figure~\ref{f3}(a) depicts transmission properties of the NIM slab at $\alpha_1L=2.3$ and different magnitudes of the absorption index $\alpha_2L>0$ that are less than, equal to and greater than $\alpha_1L$. The figure shows dramatic changes in the transmission properties with changes in the ratio of the absorption indices for the signal, $\alpha_1$, and the idler, $\alpha_2$. It is seen that the transmissions does not drop below 1 at $\alpha_2>\alpha_1$. Hence, larger absorption for the idler is advantageous for robust transmission of the signal, which is counterintuitive. The increase of the idler's absorption is followed by the relatively small shift of the resonances to larger magnitudes of $gL$.  Oscillation amplitudes grow sharply near the resonances, which indicates {cavity-less
generation}. The distribution of the signal and the idler inside the
slab would also dramatically change, as compared to that for the
first transmission minimum depicted in Fig.~\ref{f2}.  Unless
optimized, the signal maximum inside the slab may appear much greater
than its output value at $z=0$. Phase-matching of the positive- and
negative-index waves also presents a technical challenge.
Figures~\ref{f3}(b)-(d) show the possibility to significantly
{diminish the negative role of the phase mismatch on the
tailored transparency of the slab} at the expense of a modest
increase of the amplitude of the control field. At that, the spatial
distribution of the signal and the idler may experience a dramatic
change [Fig.~\ref{f2} (b)]. Such dependencies are in strong contrast
with their counterparts in PI materials and are determined by the
backwardness of the coupled waves that is inherent to NIMs.

Only rough estimations can be  made regarding $\chi^{(2)}$ attributed to metal-dielectric nanostructures. Assuming $\chi^{(2)}\sim 10^{-6}$ ESU ($\sim10^3$pm/V), which is on the order of that for CdGeAs$_2$ crystals, and a control field of $I\sim 100$ kW focused on a spot of $D\sim 50\mu$m  in diameter, one can estimate that the typical threshold value of $gL\sim1$ can be achieved for a slab thickness in the {microscopic} range of $L\sim1\mu$m, which is comparable with that of the multilayer NIM samples fabricated to date \cite{Zh,Suk}.

\subsection{Independent engineering of NLO response, quasi-resonant
four-wave mixing of ordinary and backward waves and quantum control
of transparency of NIM slab}\label{ie}
Herein, we explore the feasibility of independently engineering the
NI and the resonantly enhanced  $ \chi^{(3)}$- response of a composite metamaterial with embedded NLO
centers.  We investigate resonant and quasi-resonant FWM and
the accompanying processes that allow {coherent energy transfer}
from the control fields to the counter-propagating NI
signal and PI idler. We show that this opens up opportunities for
compensating optical losses in NIMs and for the creation of unique
NIM-based photonic microdevices, which are of critical importance for
the further development of nanophotonics in NIMs (for a review, see,
for example, \cite{Sh}). Among the specific features considered in
this study are the above-outlined {backwardness of the coupled waves and the possibility of quantum control} over the nonlinear propagation process and its outcomes. The basic scheme of resonant four-wave mixing of the backward wave in a NIM is as follows. A
slab of NIM is doped by four-level nonlinear centers [Fig.
\ref{f1}(c)] so that the signal frequency, $\omega_1$, falls in the
NI domain, whereas all other frequencies,
$\omega_3$, $\omega_4$ and $\omega_2$, are in the the PI domain. Below, we show the possibility to produce transparency
and even amplification above the oscillation threshold at $\omega_1$
controlled by two lasers at $\omega_3$ and $\omega_4$ [Fig.~
\ref{f1}(b)]. These fields generate an idler at
$\omega_2=\omega_3+\omega_4- \omega_1$, which then contributes back in
optical parametric amplification at $\omega_1$. Unlike the scheme
investigated in \cite{OLM,APB09,OL09}, here the idler corresponds to
a higher-frequency transition from the ground state, and the signal corresponds to a lower-frequency transition between the excited states. No incoherent amplification is possible here for the idler, and the dependence of the idler and the signal absorption indices on the control fields cardinally changes.  First, the scheme with relatively fast quantum coherence relaxation rates and the case where only a two-photon, Raman-like resonance for the signal holds is considered; all other one-photon frequency offsets are on the order of several tens of the optical transition widths. Then the scheme with the same quantum coherence relaxation rates but with higher partial spontaneous transition rates is considered, in which case population inversion at the coupled optical transitions is impossible. Finally we consider the scheme with longer quantum coherence lifetimes, which still does not allow population inversion at the optical transitions nor Raman-like amplification. The fact that all involved optical transitions are absorptive determines essentially different features of the overall loss-compensation technique in such composites in each proposed scheme. In all of the schemes outlined above, the linear and nonlinear local parameters can be tailored through quantum control by varying the intensities and frequency-resonance offsets for combinations of the two control driving fields.

The following model, which is characteristic of ions and some molecules embedded in a solid host, has been adopted: energy level relaxation rates $\Gamma_n=20$, $\Gamma_g=\Gamma_m=120$; partial transition probabilities  $\gamma_{gn}=50$, $\gamma_{mn}=70$, (all in $ 10^6$~s$^{-1}$); homogeneous transition half-widths $\Gamma_{lg}=1$, $\Gamma_{lm}=1.9$, $\Gamma_{ng}=1.5$, $\Gamma_{nm}=1.8$ (all in $10^{12}$~s$^{-1}$); $\Gamma_{gm}=5$, $\Gamma_{ln}=1$ (all in $10^{10}$~s$^{-1}$);  $\lambda_1=756$ nm and $\lambda_2=480$ nm. The density-matrix method \cite{GPRA} is used for calculating the intensity-dependent local parameters while accounting for the quantum nonlinear interference effects. This allows us to investigate the changes in absorption, amplification, and refractive indices as well as in the magnitudes and signs of NLO susceptibilities caused by the control fields. These changes depend on the population redistribution over the coupled levels, which in turn strongly depends on the ratio of the partial transition probabilities. Electrical linear and nonlinear polarizations, Eq. (\ref{P}),  are calculated as
\begin{eqnarray}
&&P_1(z,t)=(1/2)\{P_{01}^{L}\exp(ik_1z)\cr
&&+ P_{01}^{NL}\exp[i(k_3+k_4-k_2)z]\}
\exp(-i\omega_1 t)+ c.c.\qquad\cr
&&= N(\rho_{ng}d_{gn}+\rho_{gn}d_{ng});  \label{pe1}\\
&&P_2(z,t)=(1/2)\{P_{02}^{L}\exp(ik_2z)\cr
&&+ P_{02}^{NL}\exp[i(k_3+k_4-k_1)z]\}
\exp(-i\omega_2 t) + c.c.\qquad\cr
&&= N(\rho_{ml}d_{lm}+\rho_{lm}d_{ml}).\label{pe2}
\end{eqnarray}
Here, $\rho_{ij}$ are the density matrix elements, and $d_{ij}$ are the transition dipole elements. Effective linear, $\chi_{1,2}$, and NLO, $\chi_{1,2}^{(3)}$, susceptibilities dependent on the intensities of the
driving control fields $E_3$ and $E_2$ are defined as
\begin{eqnarray}
P_{01}^{L}&=&\chi_1E_1,\quad P_{01}^{NL}=\chi_1^{(3)}E_3E_4E_2^{\ast};\\ \label{ch1}
P_{02}^{L}&=&\chi_2E_2,\quad  P_{02}^{NL}=\chi_2^{(3)}E_3E_4E_1^{\ast}. \label{ch2}
\end{eqnarray}
The linear susceptibilities determine the intensity-dependent contributions to absorption and to the refractive indices of the composite attributed to the embedded centers, while the NLO susceptibilities determine the FWM. Here, $\omega_{1}+\omega_{2}=\omega_{3}+\omega_{4}$, and $k_{j}=|n_{j}|\omega _{j}/c>0$.
\begin{figure}[!h]
\begin{center}
\includegraphics[width=.454\columnwidth]{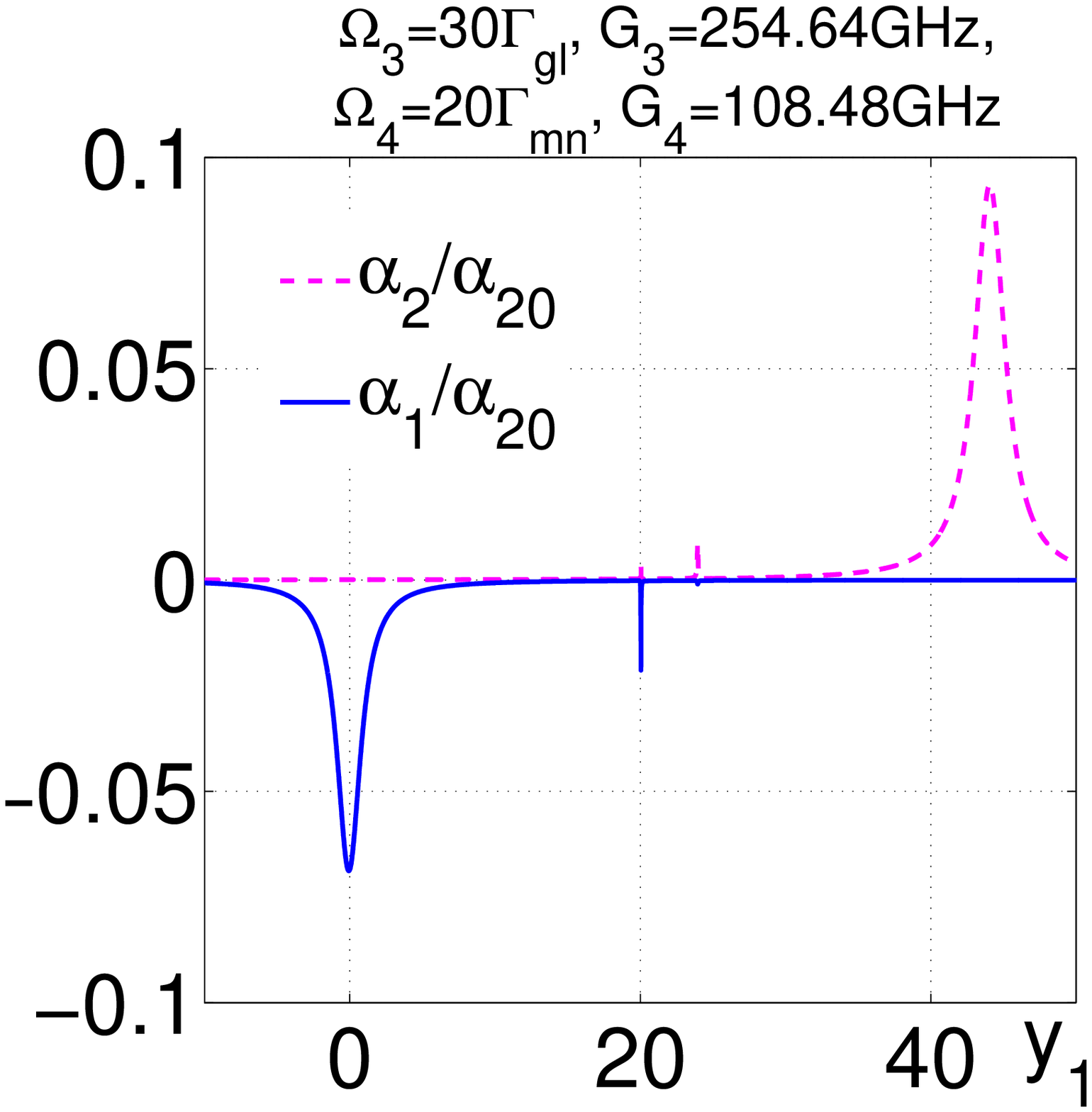}
\includegraphics[width=.454\columnwidth]{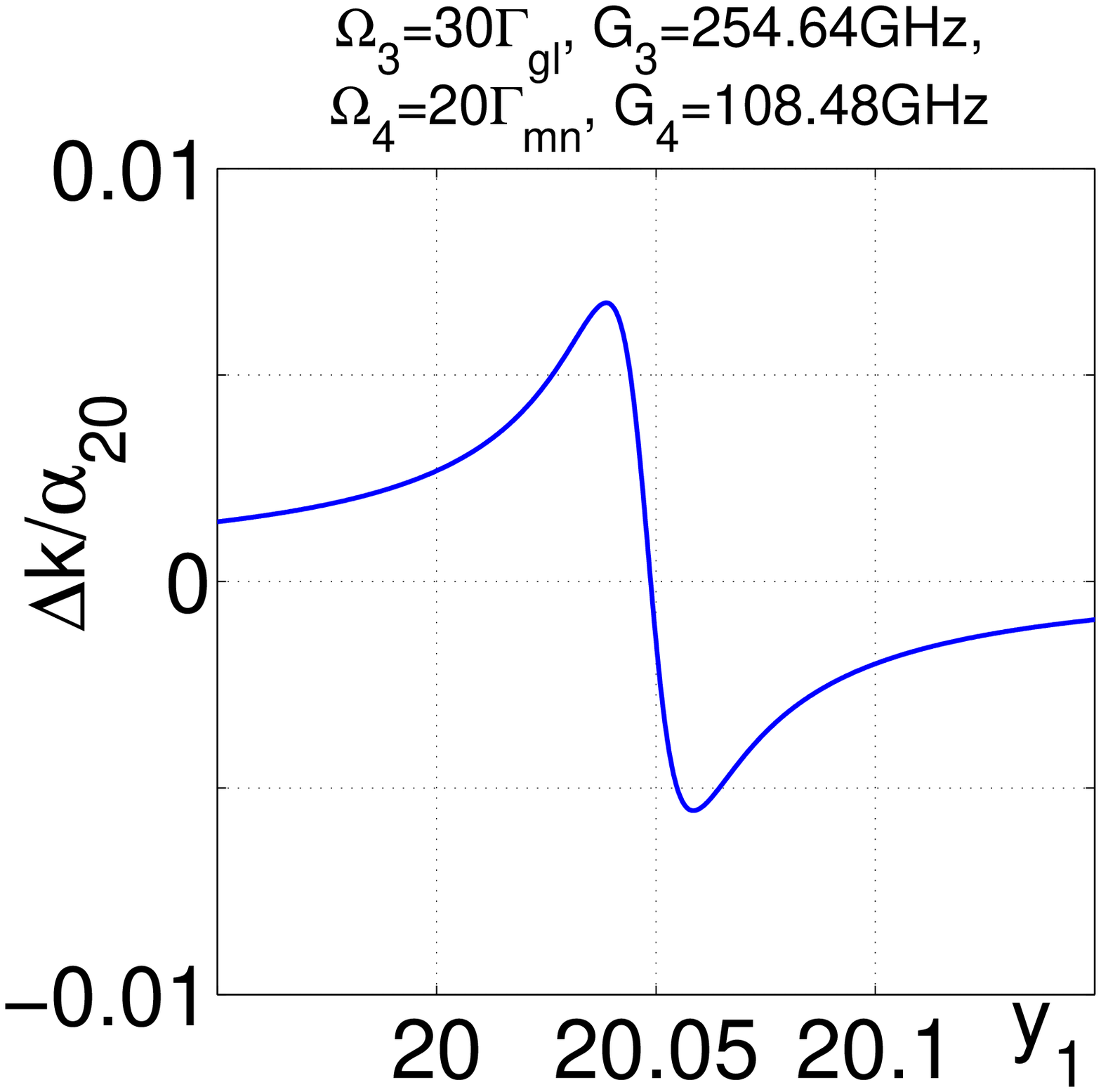}\\
(a)\hspace{30mm} (b)\\
\includegraphics[width=.454\columnwidth]{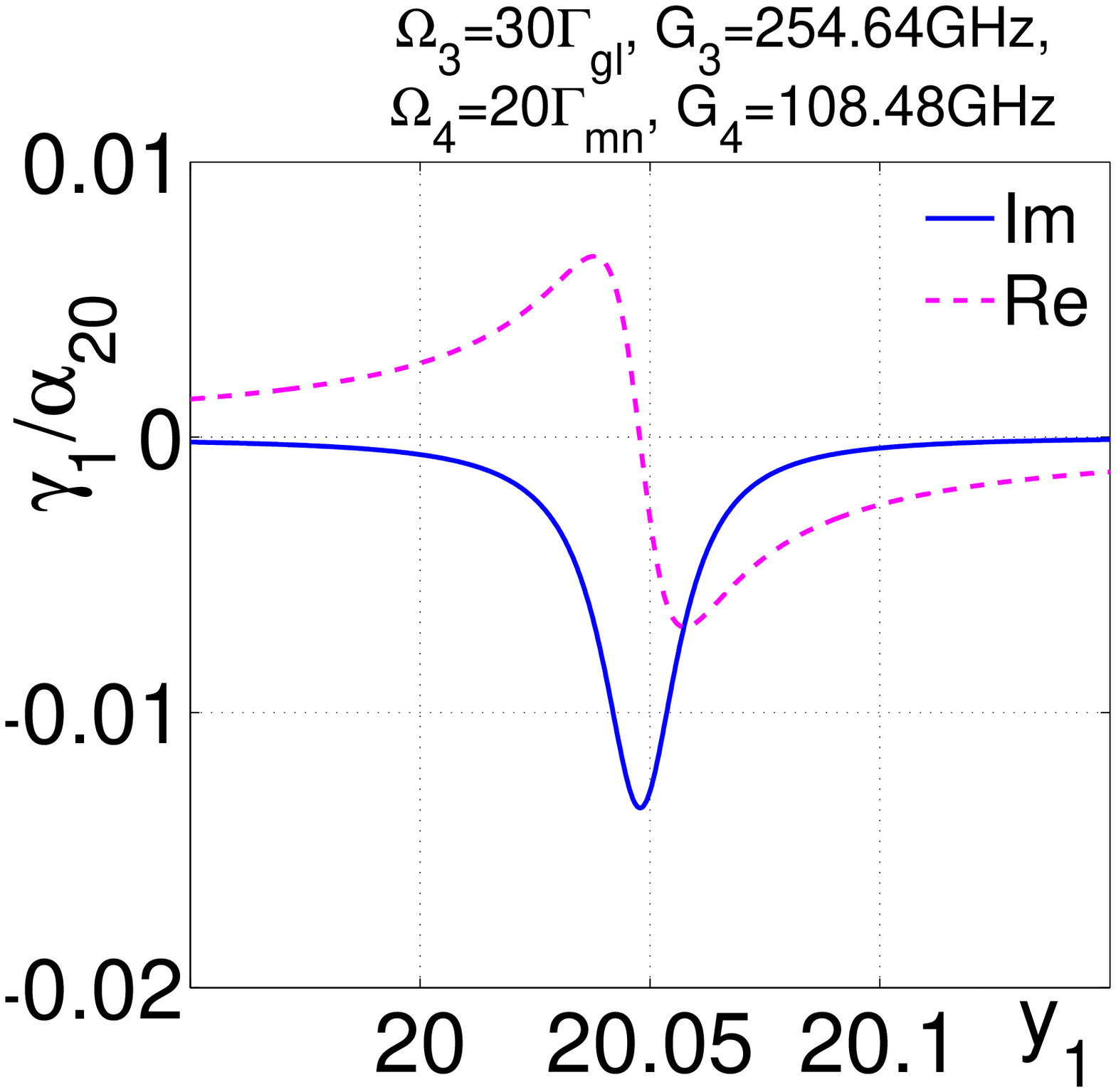}
\includegraphics[width=.454\columnwidth]{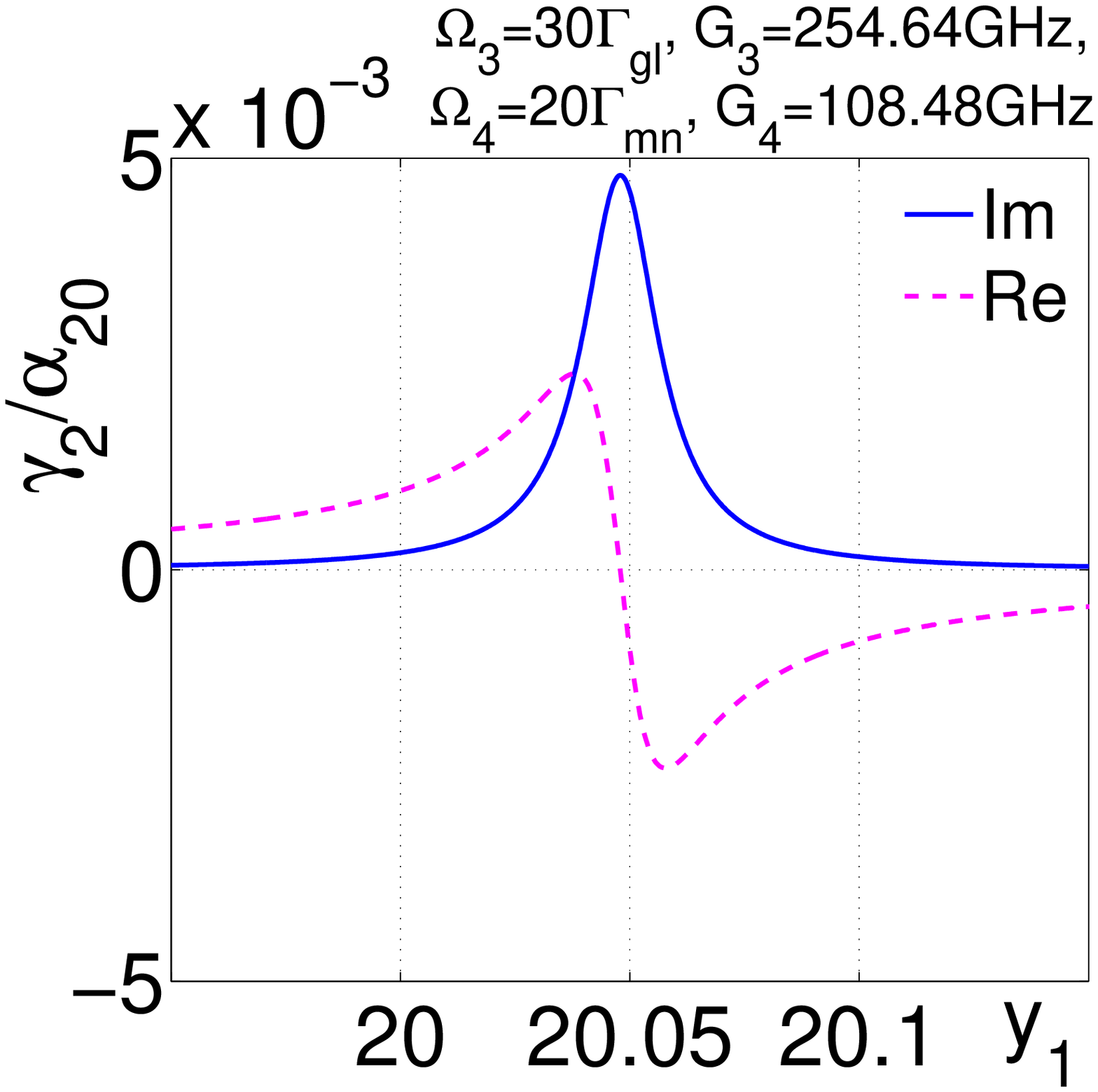}\\
(c)\hspace{30mm} (d)\\
\includegraphics[width=.454\columnwidth]{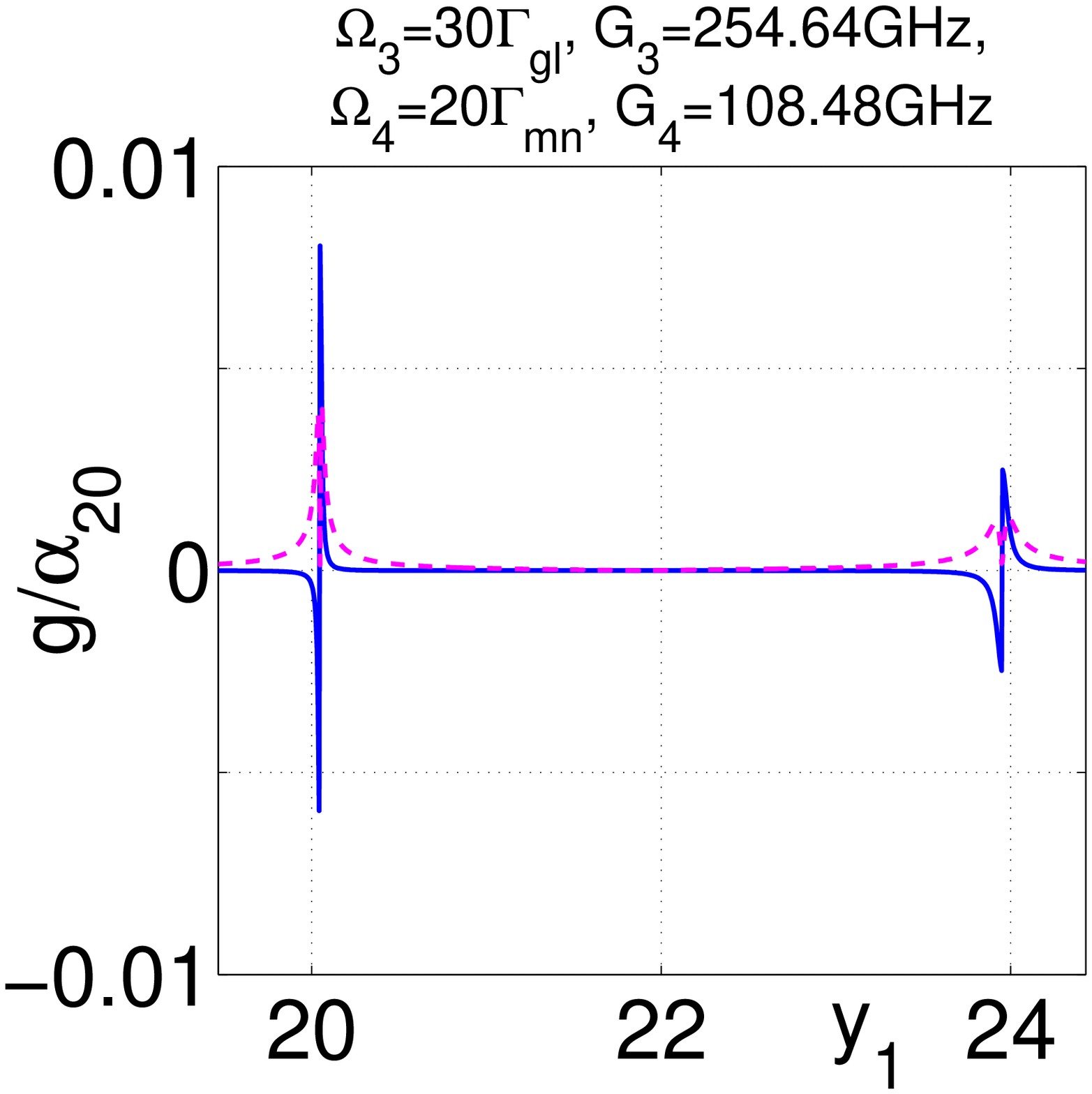}
\includegraphics[width=.454\columnwidth]{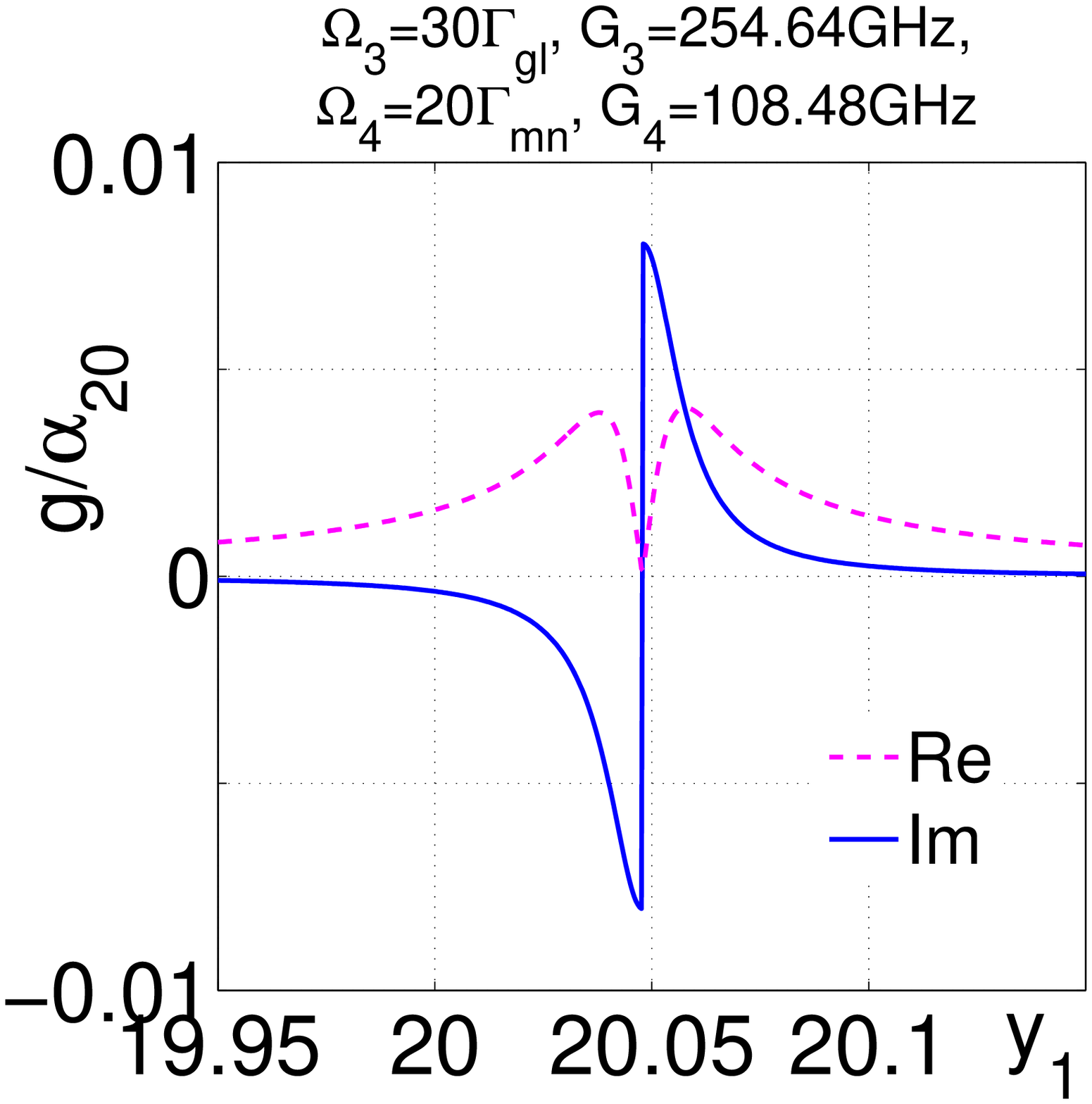}\\
(e)\hspace{30mm} (f)
\end{center}
\caption{Nonlinear spectral structures in local optical quantities produced by the control fields. $y_1=(\omega_1-\omega_{gn})/\Gamma_{gn}$, $\omega_2=\omega_3+\omega_4-\omega_1$. (a): absorption/gain indices for the
signal and the idler; (b): phase mismatch; (c)-(f): four-wave mixing
coupling parameters.  Coupling
Rabi frequencies and resonance frequency offsets for the control
fields are: $G_3=254.64$ GHz, $\Omega_3=30\Gamma_{gl}$,
$G_4=108.48$ GHz, $\Omega_4=20 \Gamma_{mn}$.  } \label{f4}
\end{figure}
\begin{figure}[!h]
\begin{center}
\includegraphics[width=.4\columnwidth]{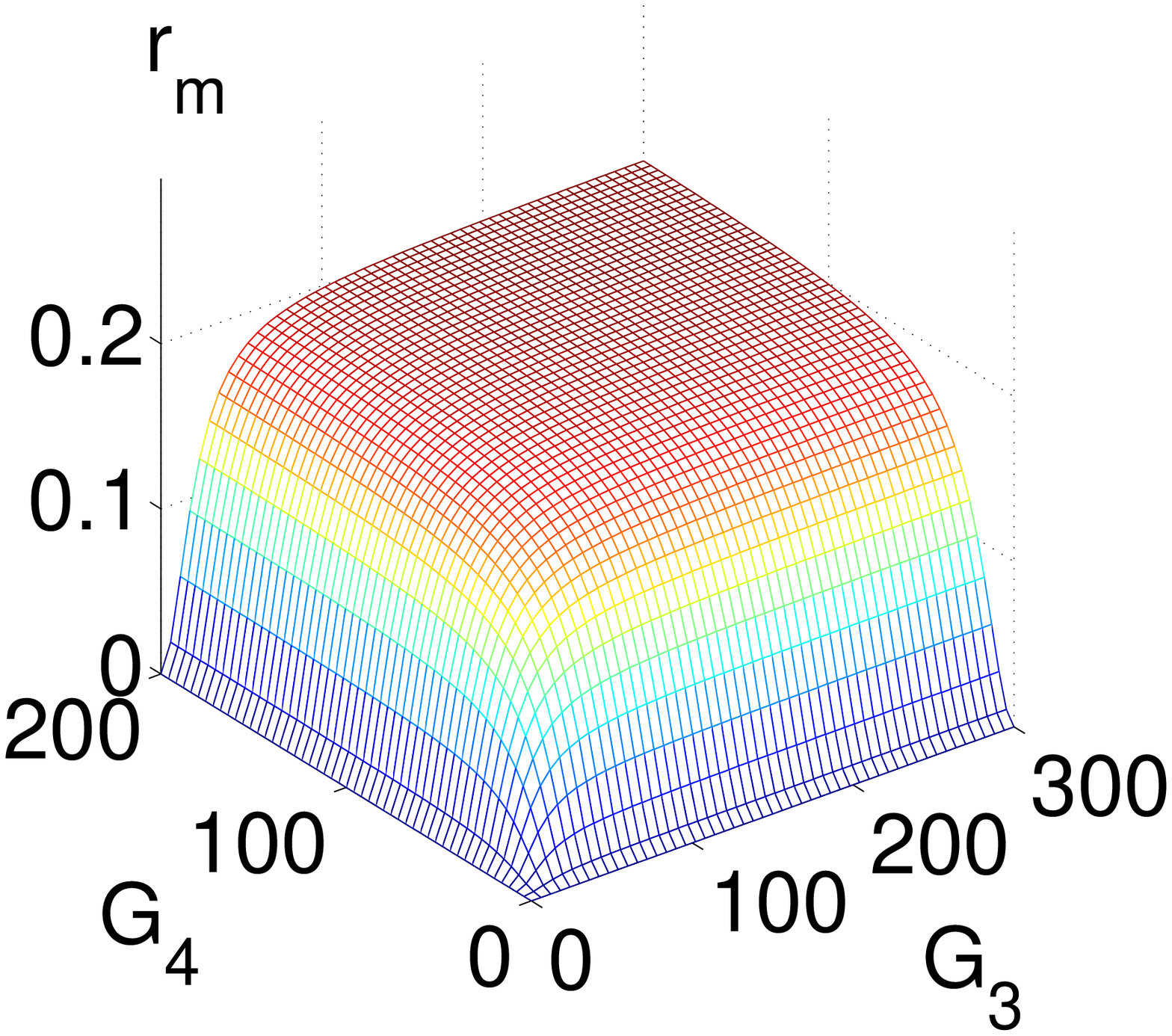}
\includegraphics[width=.4\columnwidth]{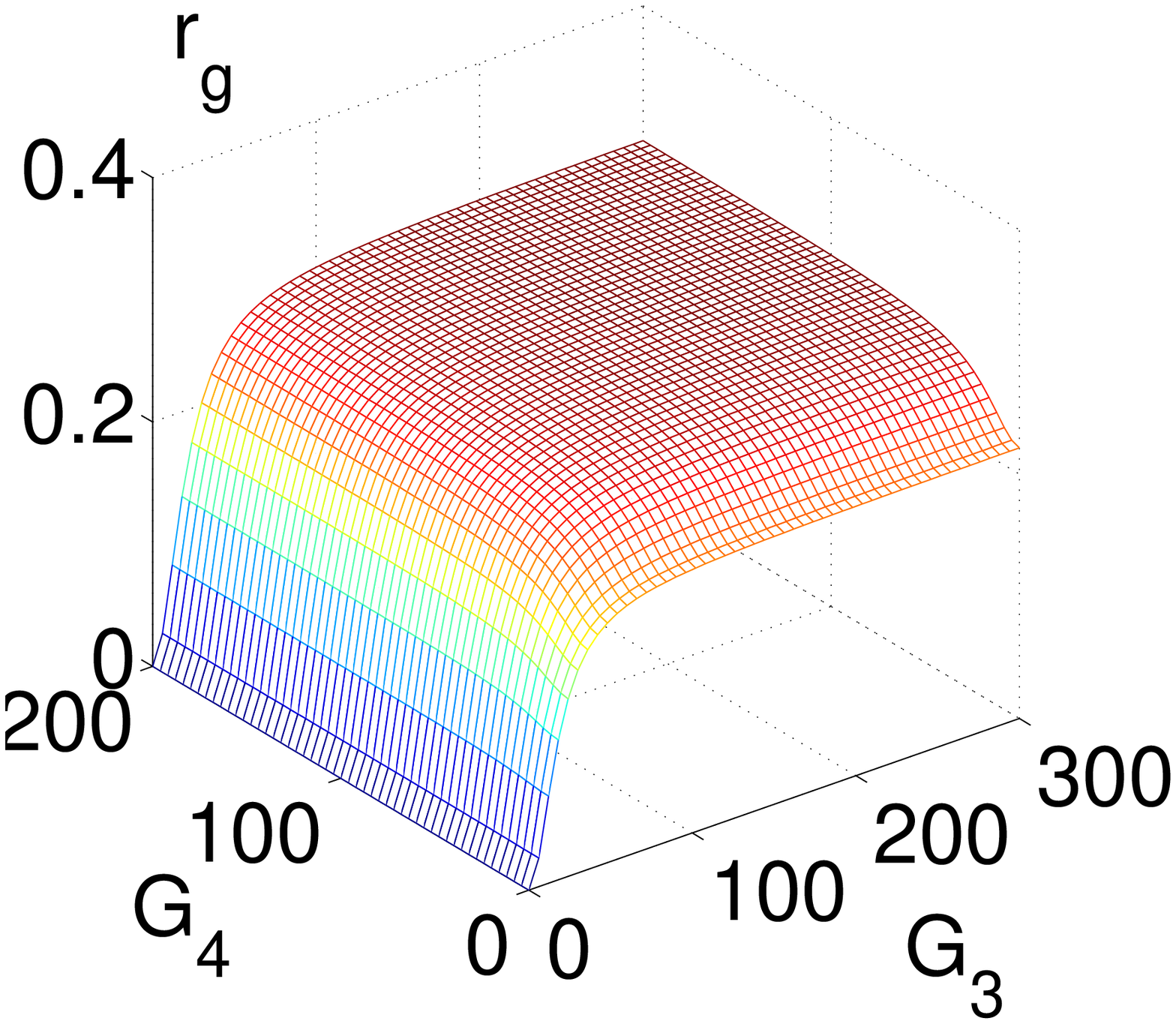}\\
(a)\hspace{30mm} (b)\\
\includegraphics[width=.4\columnwidth]{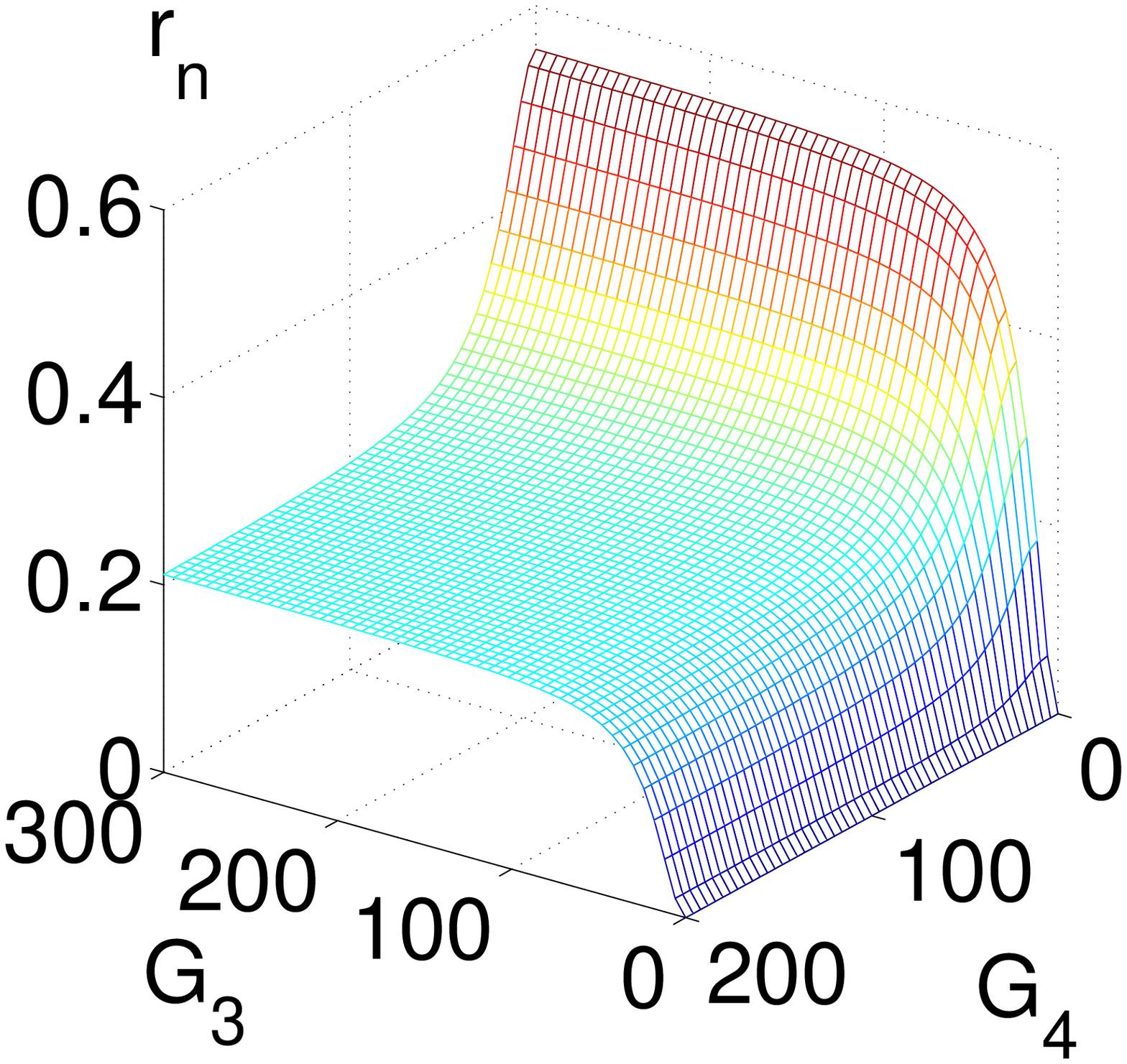}
\includegraphics[width=.4\columnwidth]{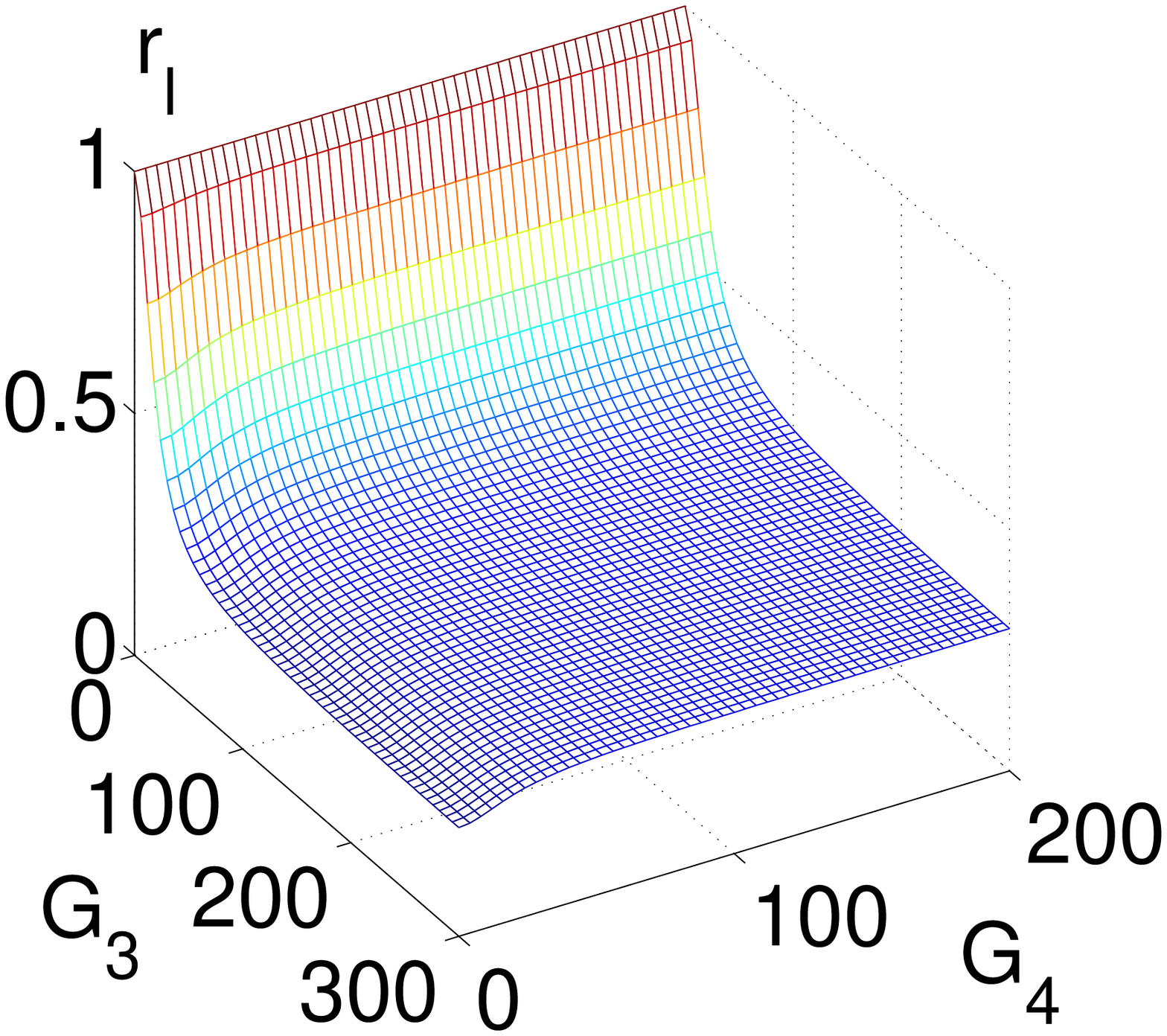}\\
(c)\hspace{30mm} (d)\\
\includegraphics[width=.31\columnwidth]{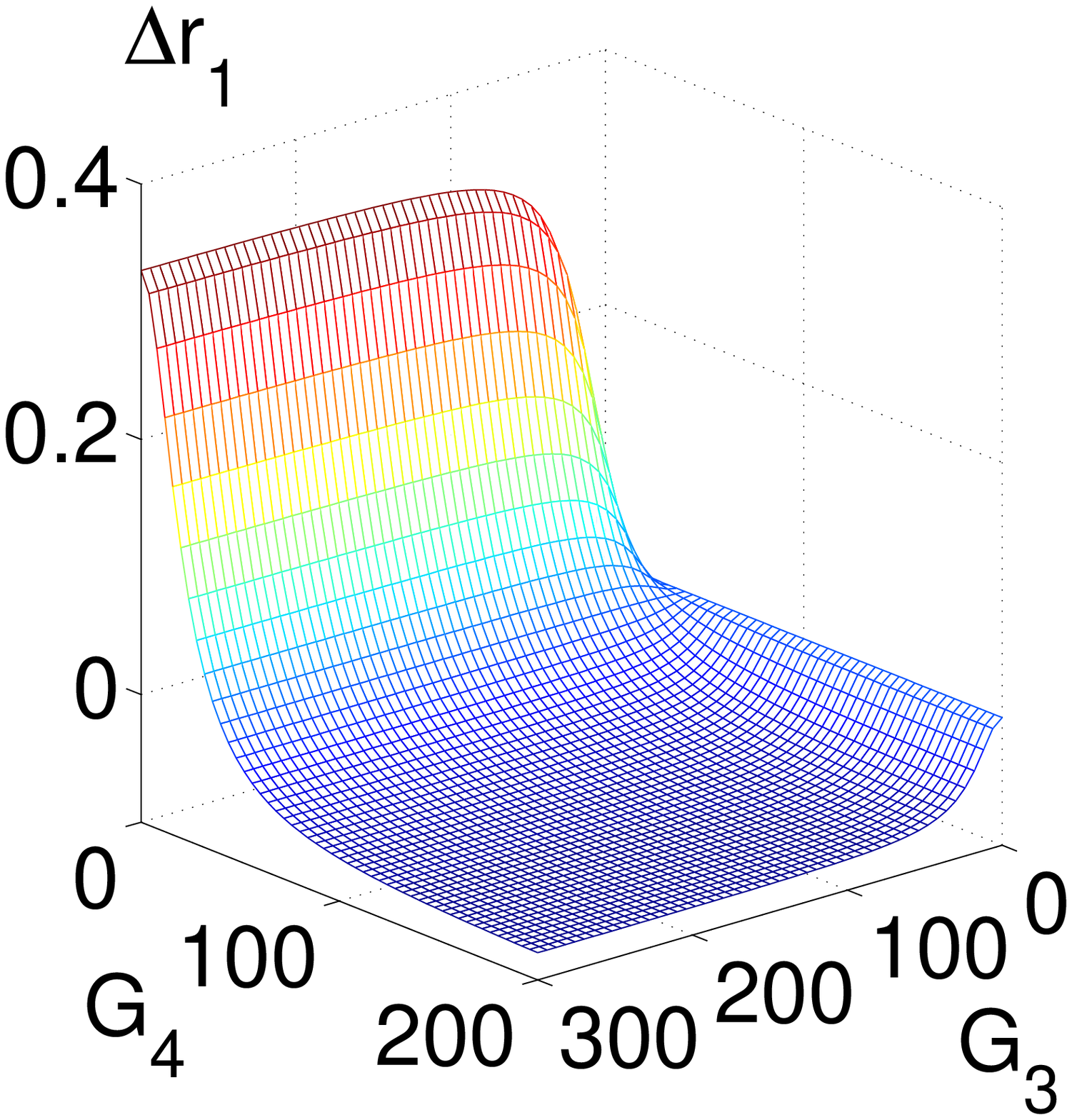}
\includegraphics[width=.31\columnwidth]{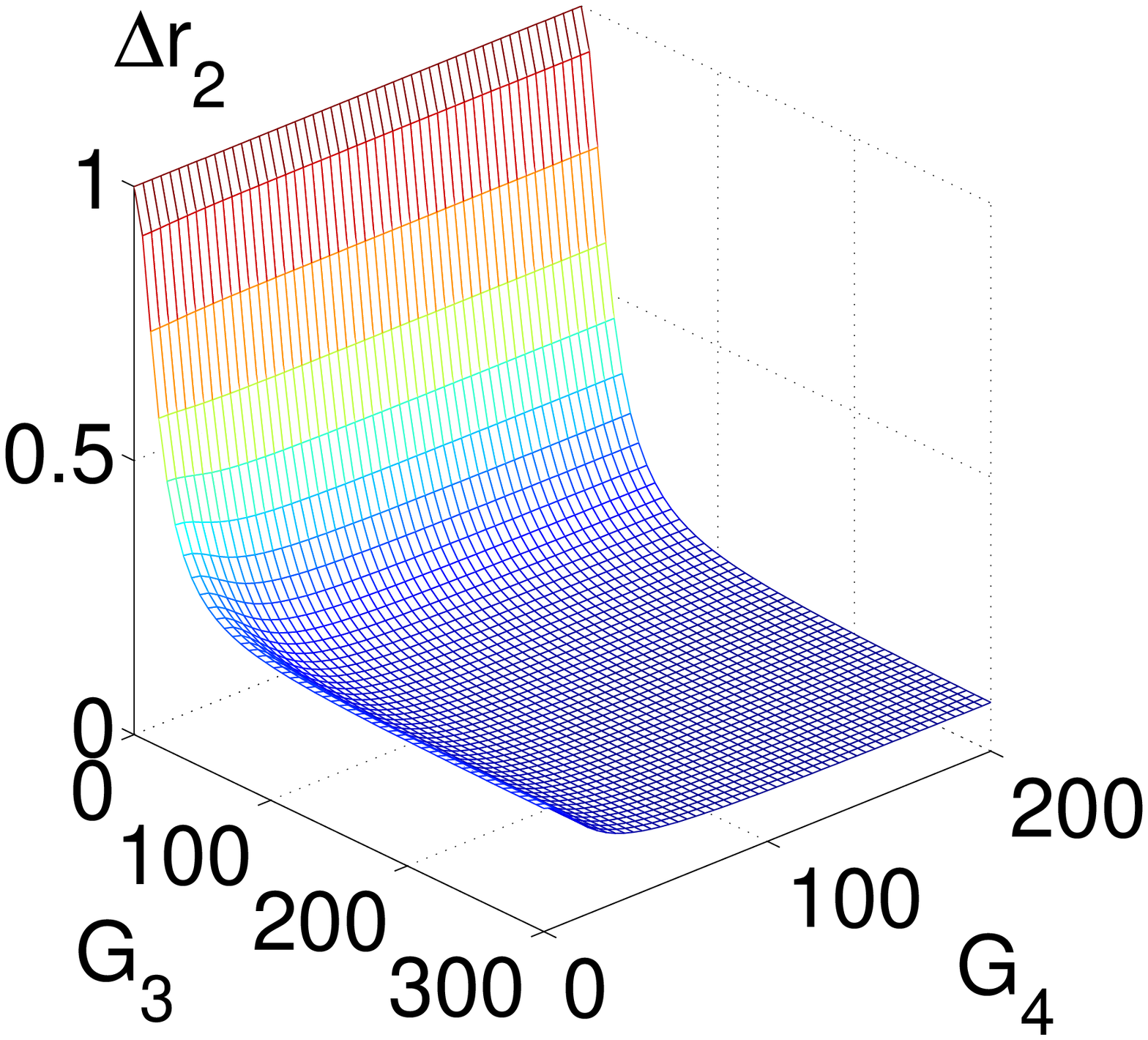 }
\includegraphics[width=.31\columnwidth]{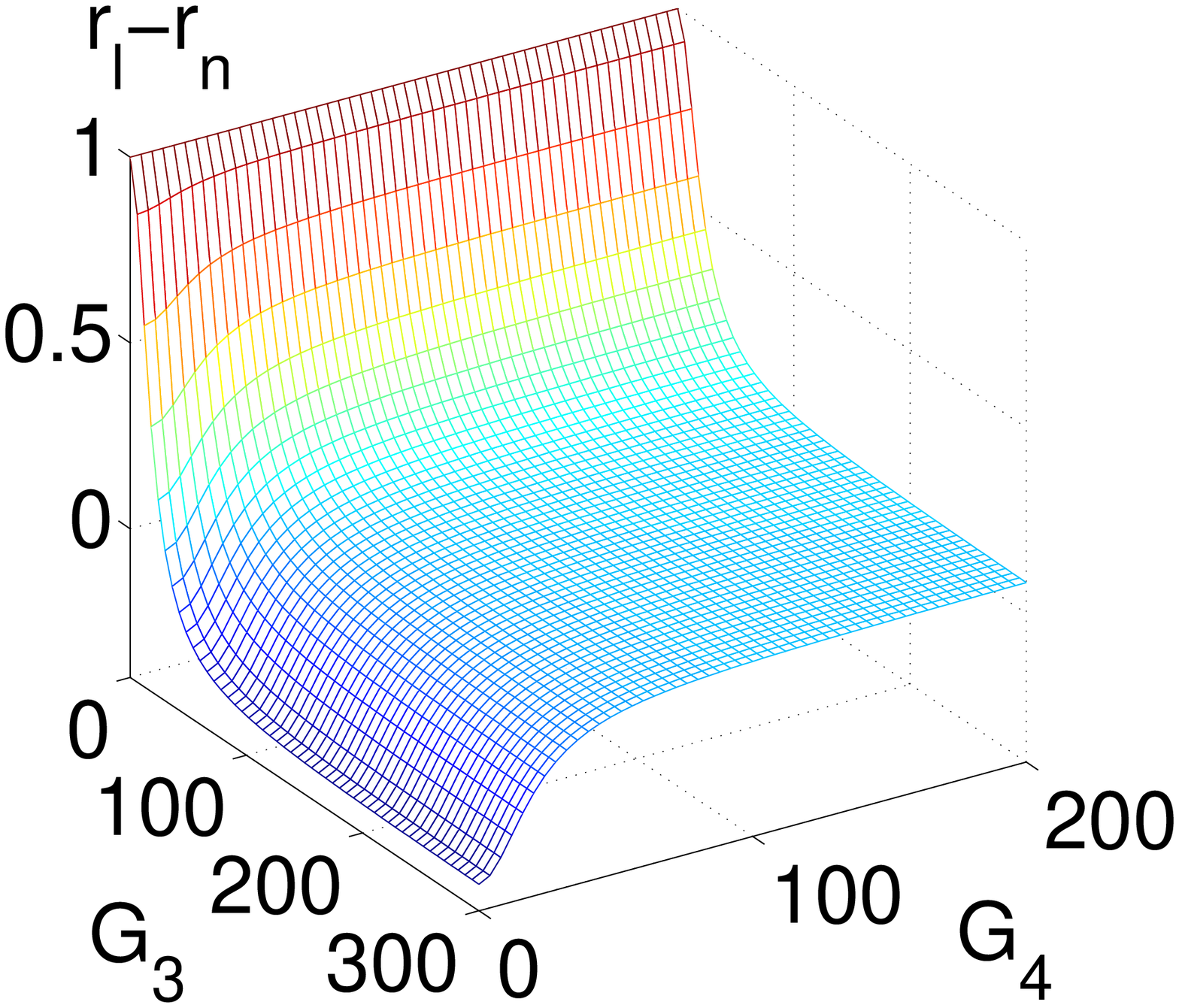 }\\
(e)\hspace{20mm} (f)\hspace{20mm} (g)\\
\includegraphics[width=.454\columnwidth]{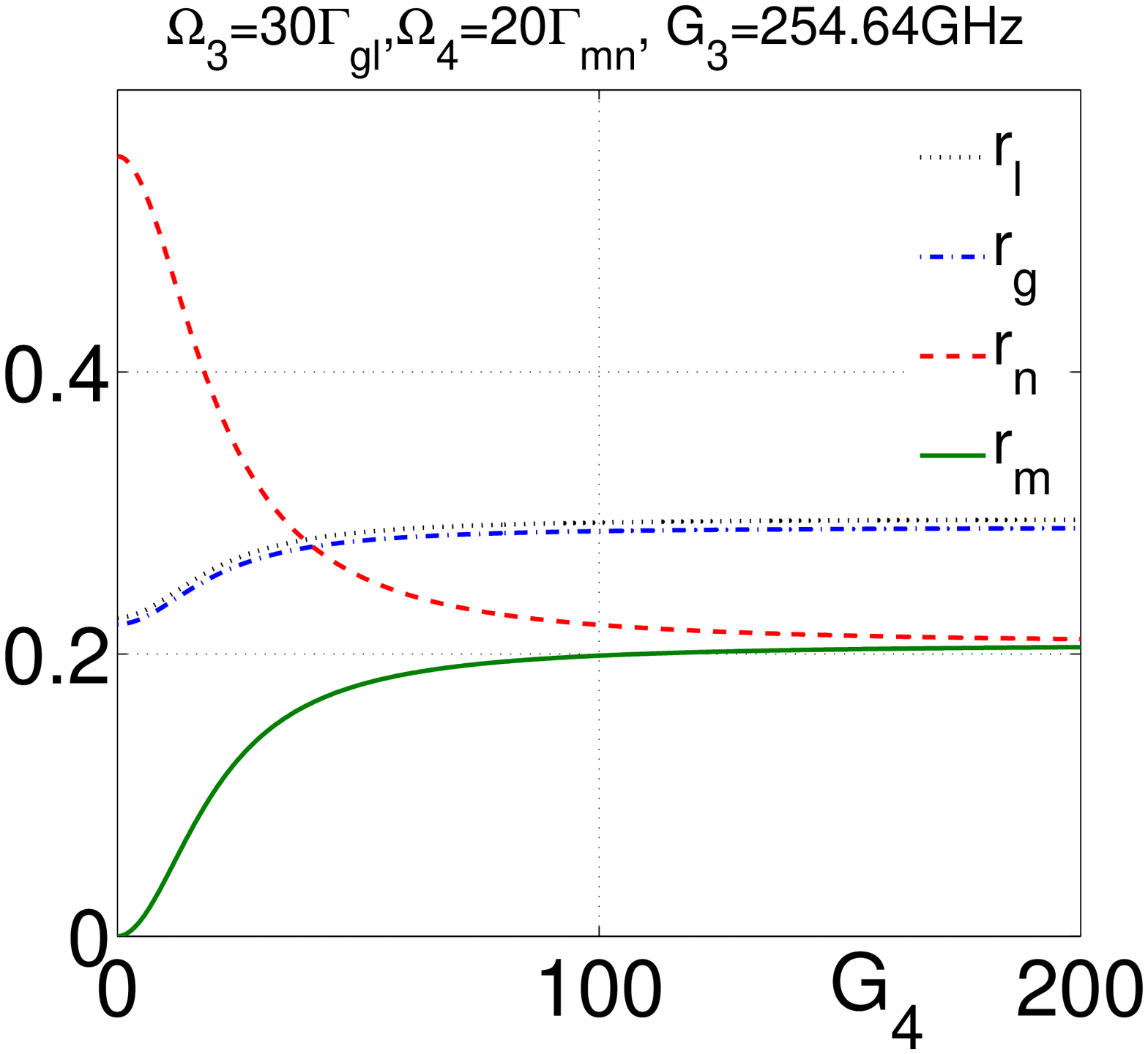}
\includegraphics[width=.454\columnwidth]{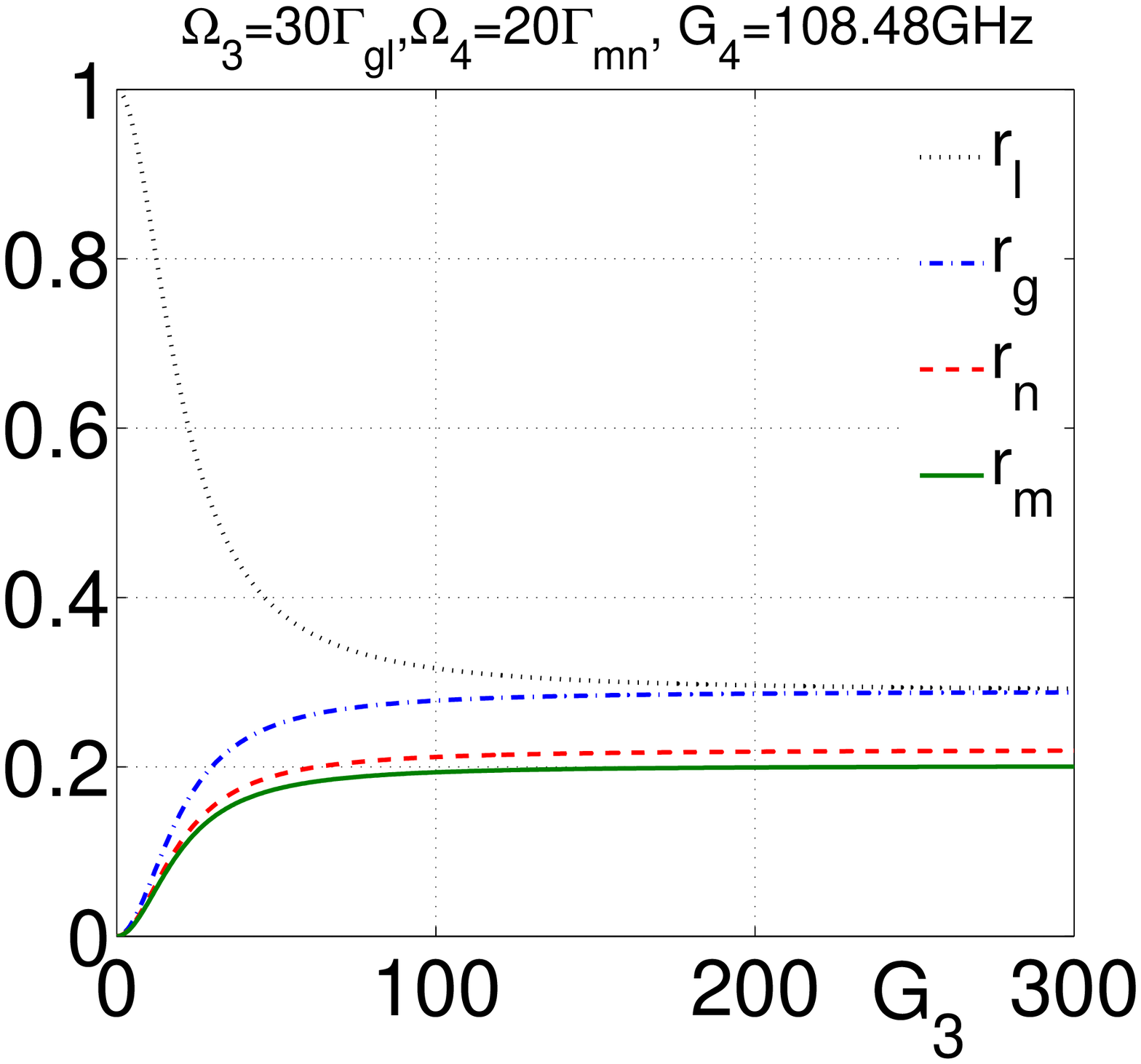}\\
(h)\hspace{30mm} (i)
\end{center}
\caption{Difference of the energy-level populations and their
dependence on the Rabi frequency of the control fields $G_3$ and $G_4$ (given in GHz).
$\Omega_3=30 \Gamma_{gl}$, $\Omega_4=20 \Gamma_{mn}$. (h): $G_3$=254.64 GHz, (i): $G_4$=108.48 GHz.} \label{f5}
\end{figure}
A {significant difference} between the resonant and off-resonant NLO processes is that all local optical parameters become intensity-dependent, and hence their spectral properties may experience a radical change near resonance. In particular, the NLO susceptibilities and, therefore, the parameters $\gamma_1$ and $\gamma_2$ become complex and differ from each other in the vicinity of the resonances. Hence, the factor $g^2$ may become negative or complex. This indicates an additional
phase shift between the NLO polarization and the generated wave
that causes {further radical changes in the nonlinear
propagation features}, which can be tailored.

\begin{figure}[!h]
\begin{center}
\includegraphics[width=.454\columnwidth]{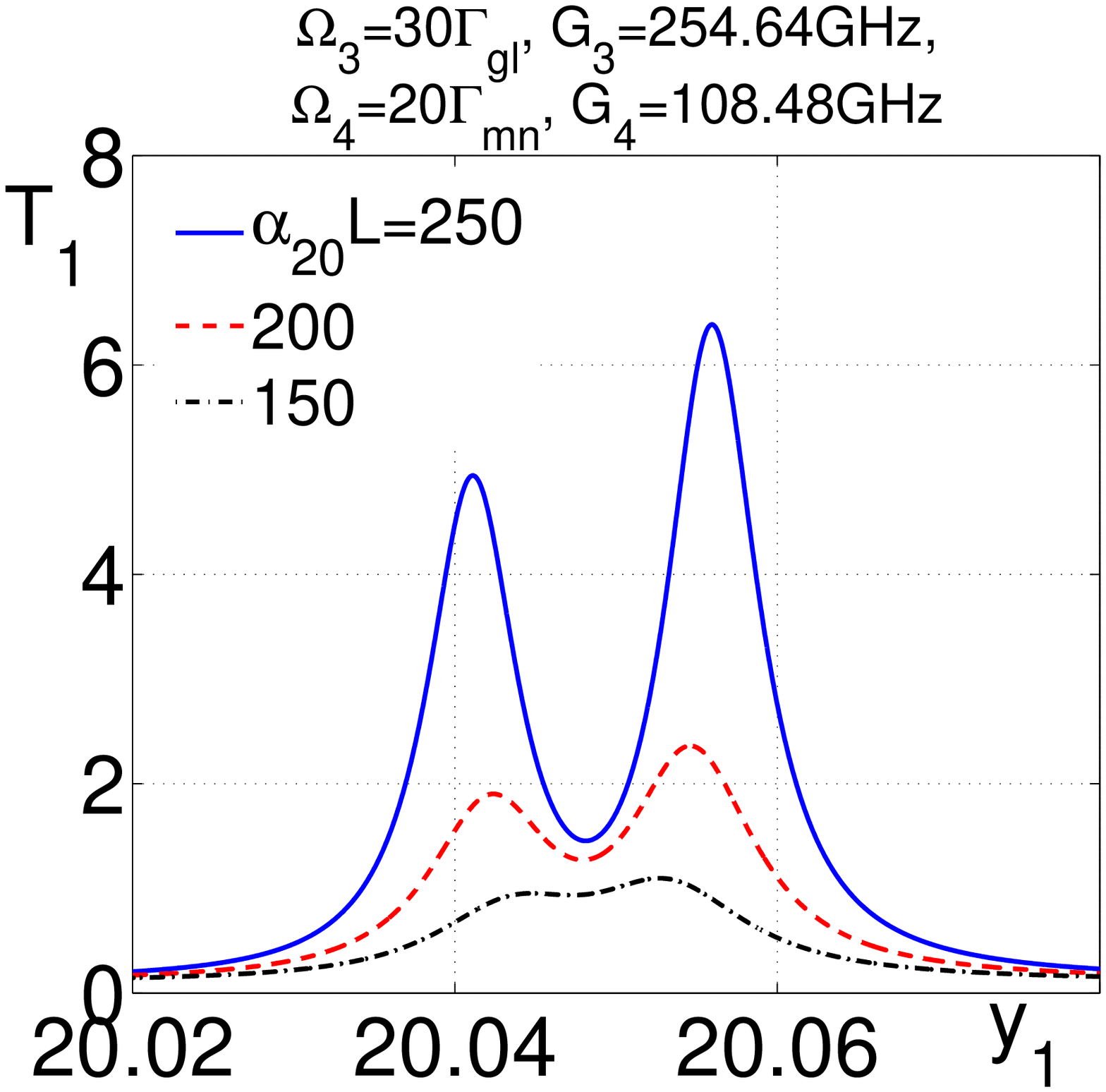}
\includegraphics[width=.454\columnwidth]{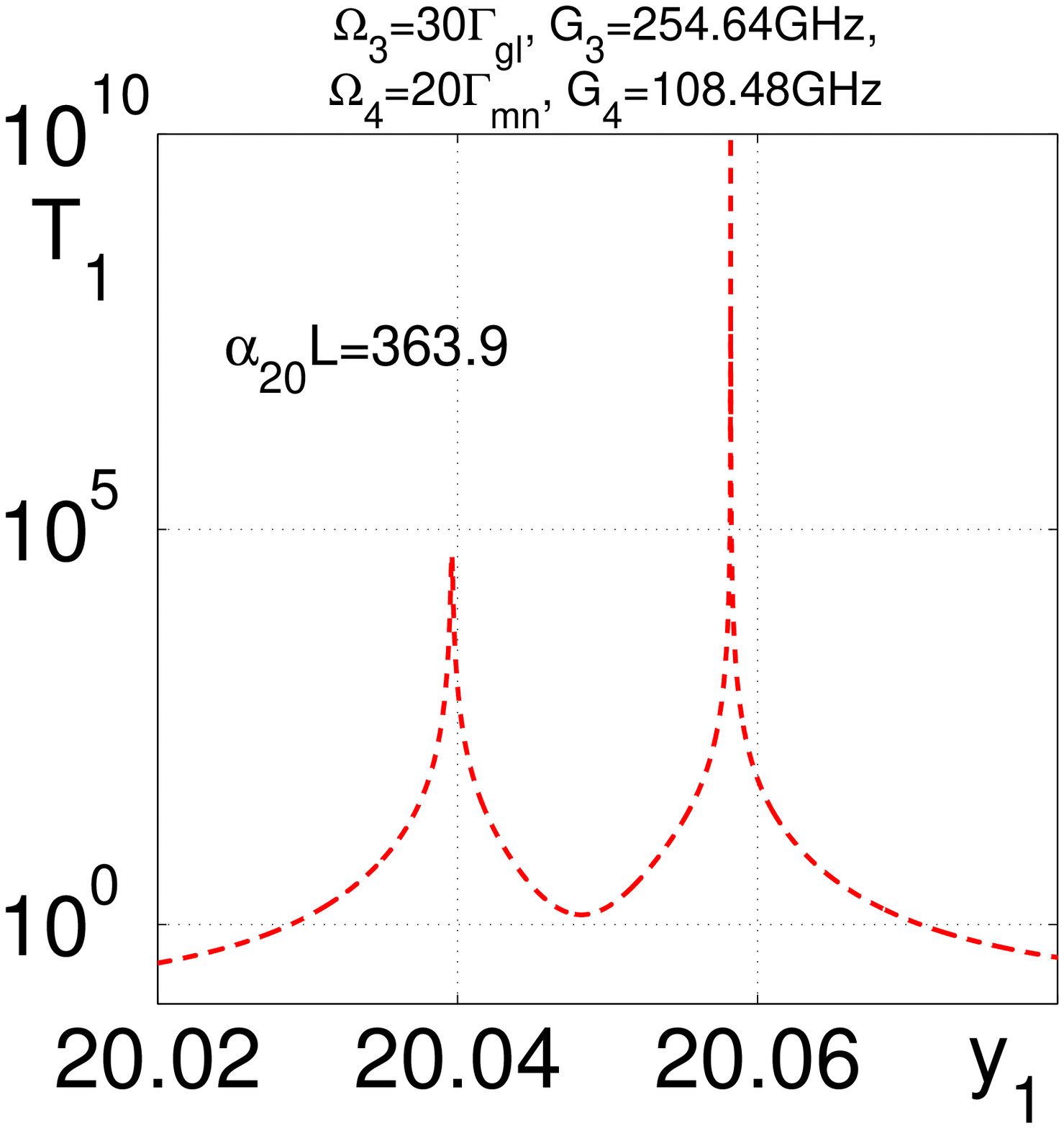}\\
(a)\hspace{30mm} (b)\\
\includegraphics[width=.454\columnwidth]{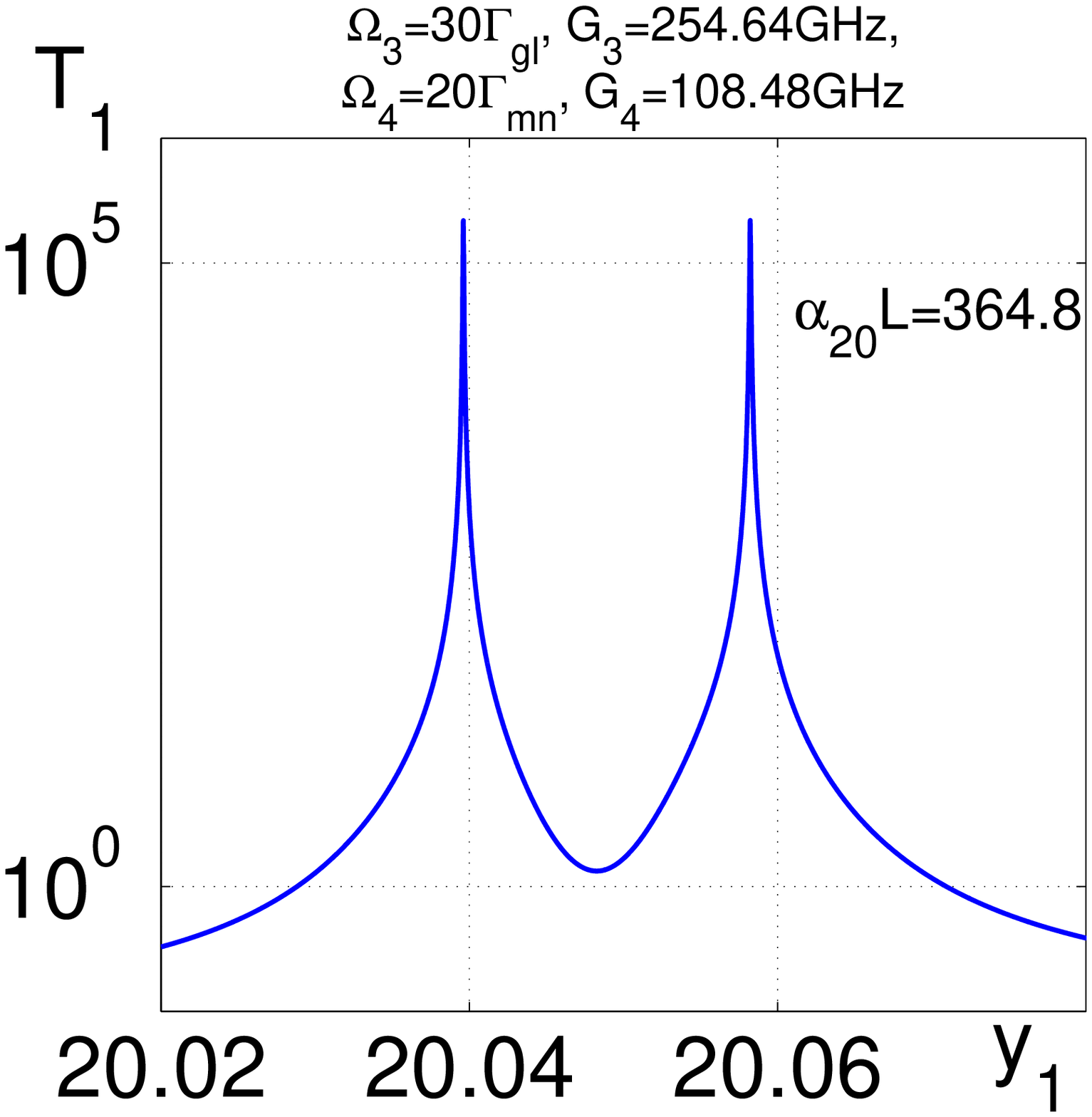}
\includegraphics[width=.454\columnwidth]{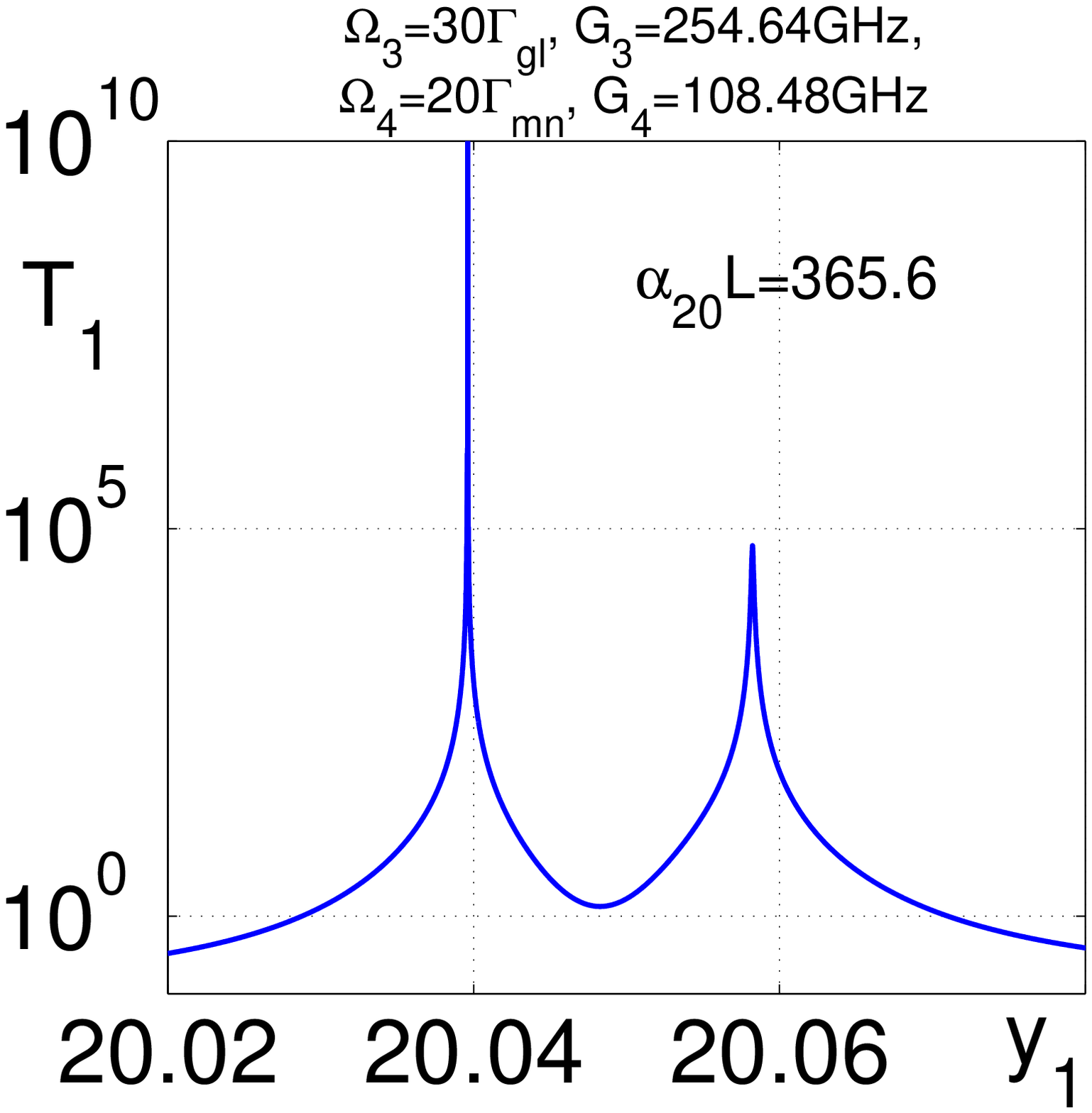}\\
(c)\hspace{30mm} (d)\\
\includegraphics[width=.454\columnwidth]{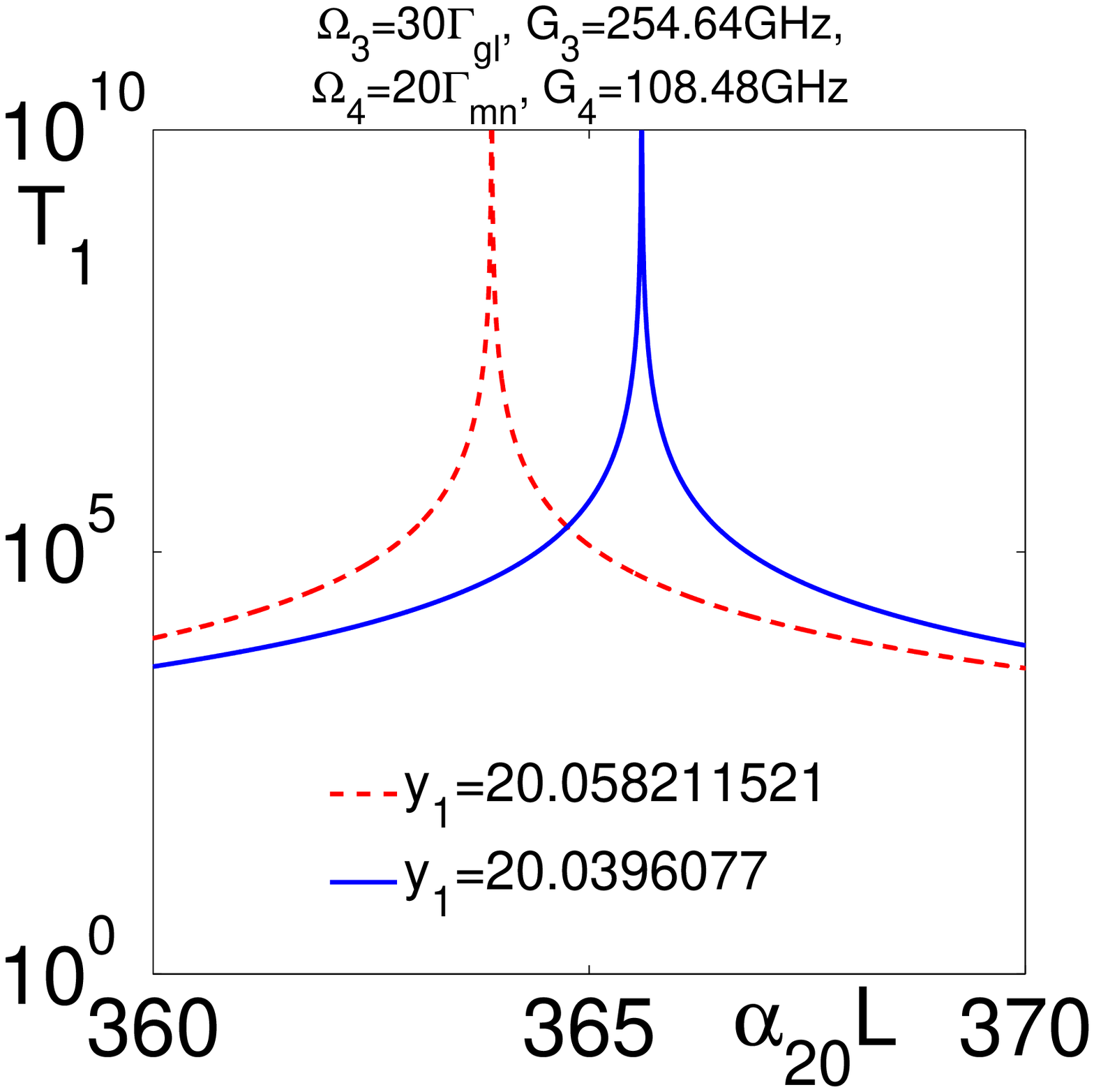}
\includegraphics[width=.454\columnwidth]{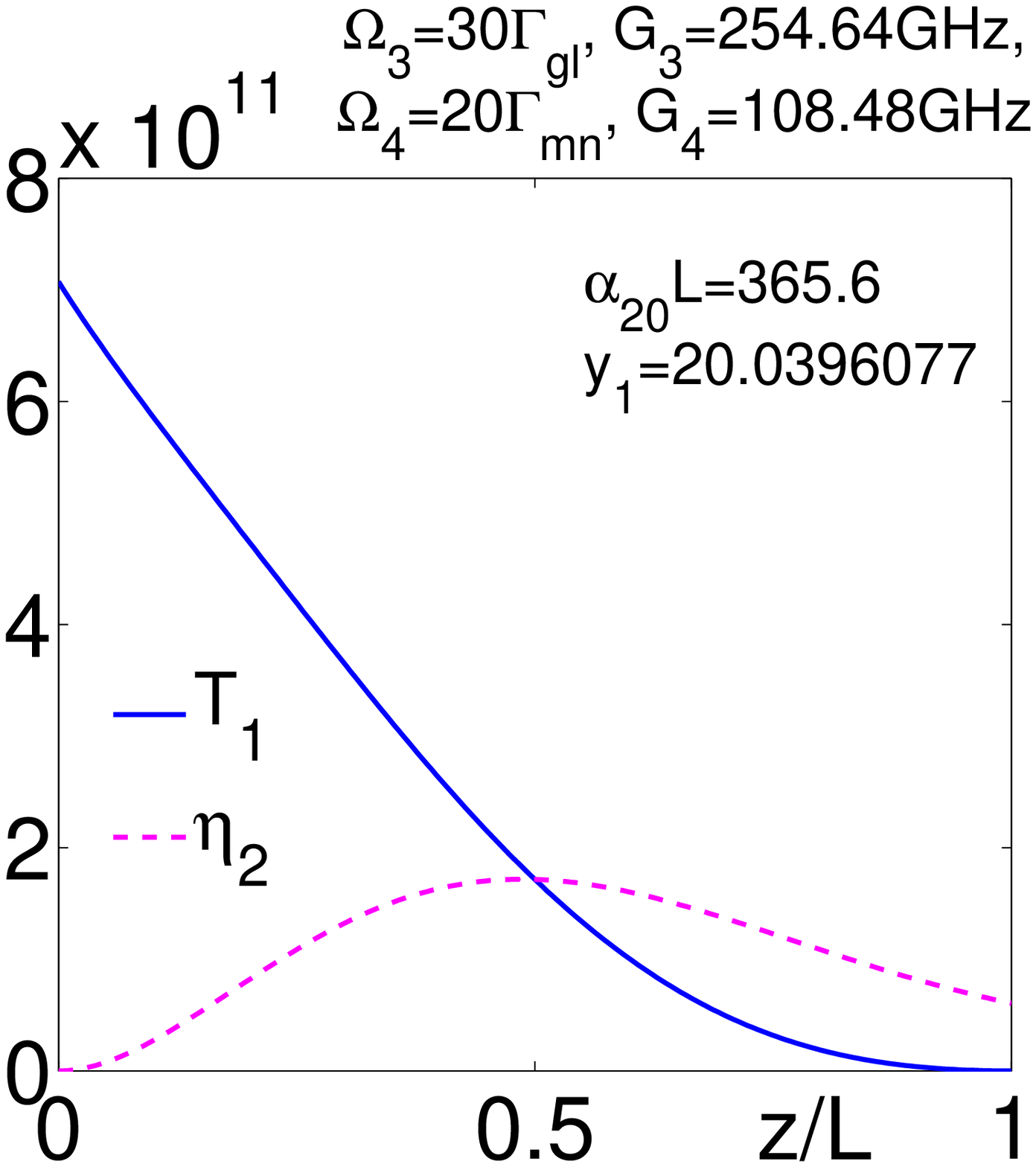}\\
(e)\hspace{30mm} (f)
\end{center}
\caption{Dependence of the transmission of the slab on the resonance
frequency offset $y_1=(\omega_1-\omega_{gn})/\Gamma_{gn}$ for
different optical densities of the slab, (a)-(d), on the resonant optical density of the slab, (e), and the distribution of the signal and the idler along the slab, (f). $G_3$=254.64 GHz, $\Omega_3= 30 \Gamma_{gl}$,
$G_4$=108.48 GHz, $\Omega_4= 20 \Gamma_{mn}$. } \label{f6}
\end{figure}
Figure~\ref{f4} depicts such modifications at the given resonance
offsets and intensities of the control fields.  Here,
$\Omega_1=\omega_1-\omega_{gn}$; other resonance detunings
$\Omega_{j}$ are defined in a similar way. Coupling Rabi frequencies
are introduced as $G_{3}= E_3d_{lg}/2\hbar$ and $G_{4} =
E_4d_{nm}/2\hbar$. The quantity  $\alpha_{20}$ denotes the fully
resonant value of absorption introduced by the embedded centers at
$\omega_2=\omega_{ml}$ with all driving fields turned off.
Figure~\ref{f4}(a) displays the modified absorption/gain indices.
The nonlinear spectral structures are caused by the
modulation of the probability amplitudes, which exhibits itself as
an effective splitting of the energy levels coupled with the driving
fields.  Figure~\ref{f4}(b) shows the contribution to the phase
mismatch associated with one such spectral structure.
Figure~\ref{f4}(c) and (d) indicate that the real and imaginary
parts of the NLO susceptibilities become commensurate for the
given susceptibility, but may exceed their counterparts for the idler by several times. This occurs due to the fact that
different population differences contribute in different ways to
the NLO susceptibilities \cite{GPRA}, and driving fields cause
significant redistributions of the level populations (Fig.
\ref{f5}). At the given partial probabilities of spontaneous transition
between the levels, population inversions at the signal transition become possible [Fig.\ref{f5}(e),(g)-(i)]. However, for the given frequency offsets of the control fields, corresponding amplification contributes negligibly to the energy conversion [Fig.\ref{f4}(a)]. Alternatively, two-photon, Raman-like amplification at $\Omega_1\approx 20.05 \Gamma_{gn}$ shown in  Fig.\ref{f4}(a) supports coherent, parametric energy-conversion from the control fields to the signal.
\begin{figure}[!h]
\begin{center}
\includegraphics[width=.4\columnwidth]{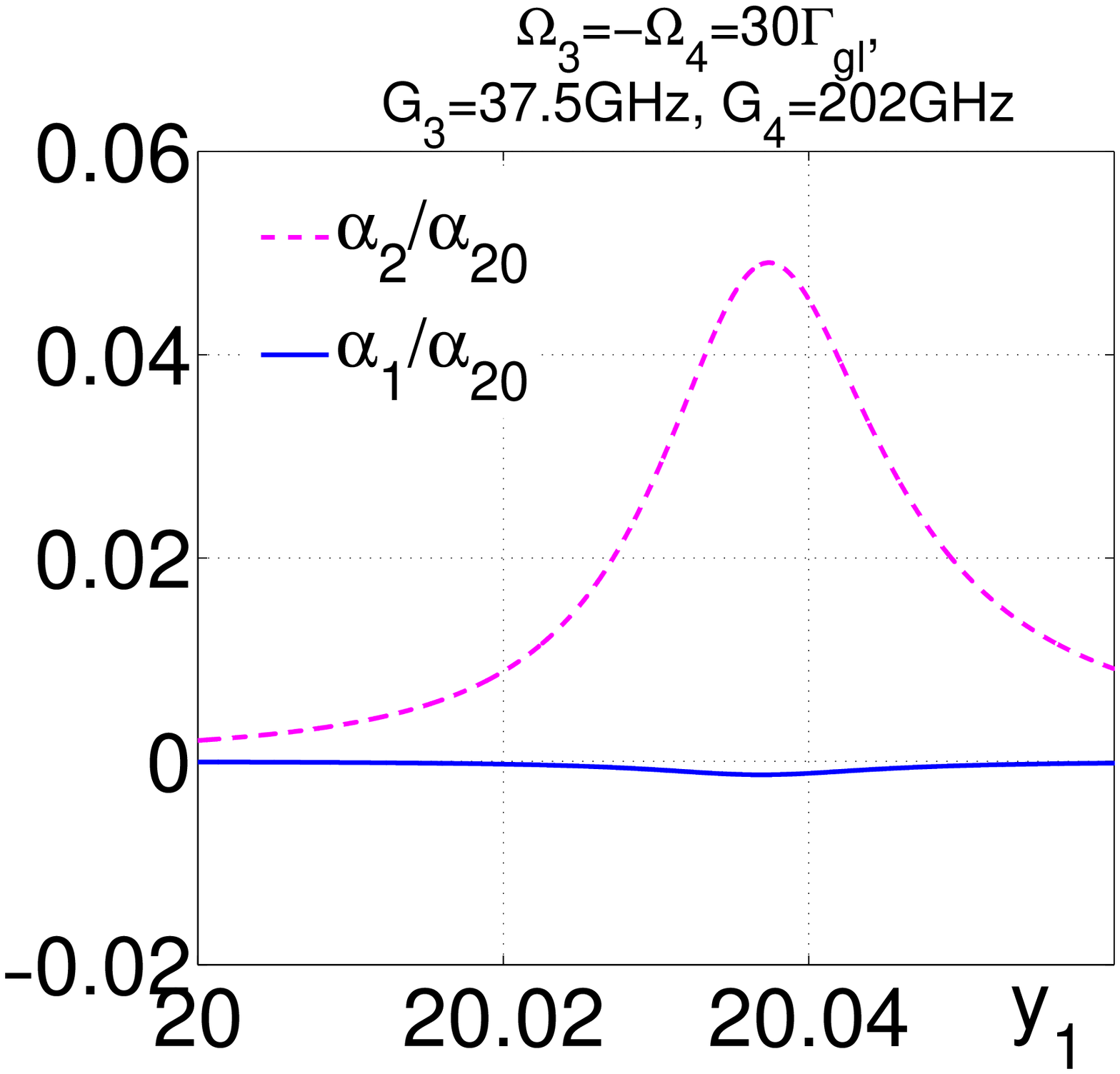}
\includegraphics[width=.4\columnwidth]{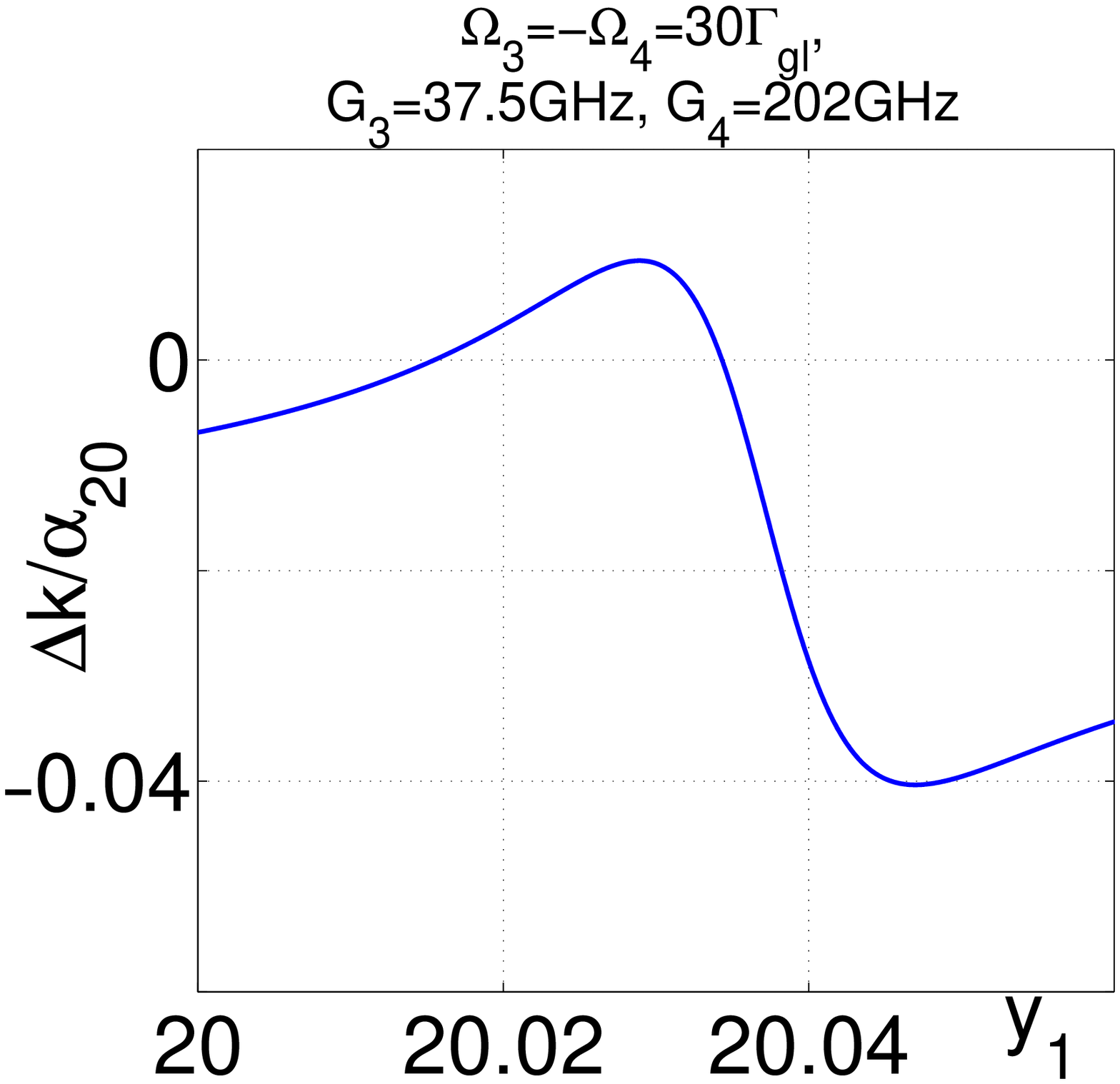}\\
(a)\hspace{30mm} (b)\\
\includegraphics[width=.4\columnwidth]{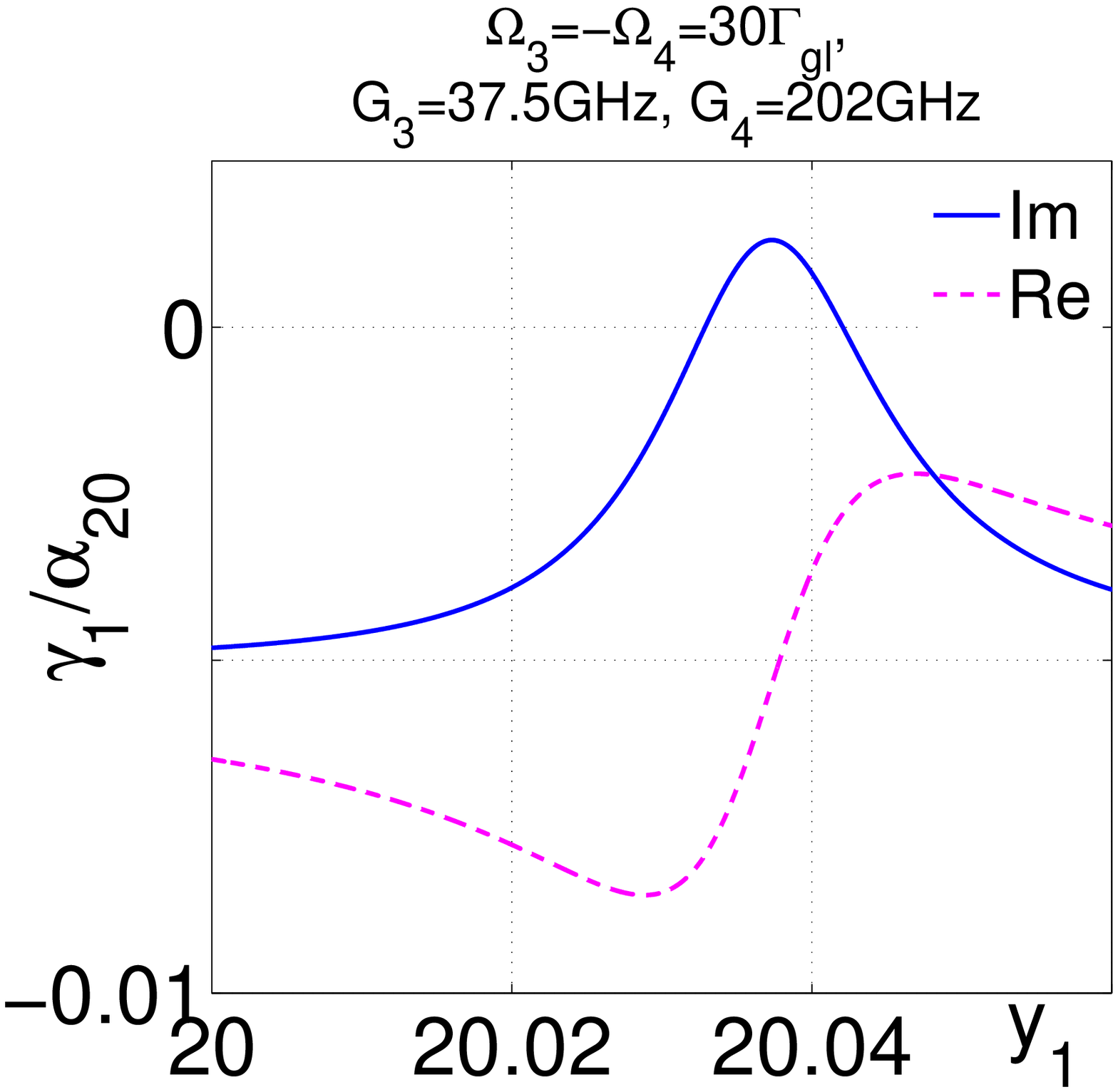}
\includegraphics[width=.4\columnwidth]{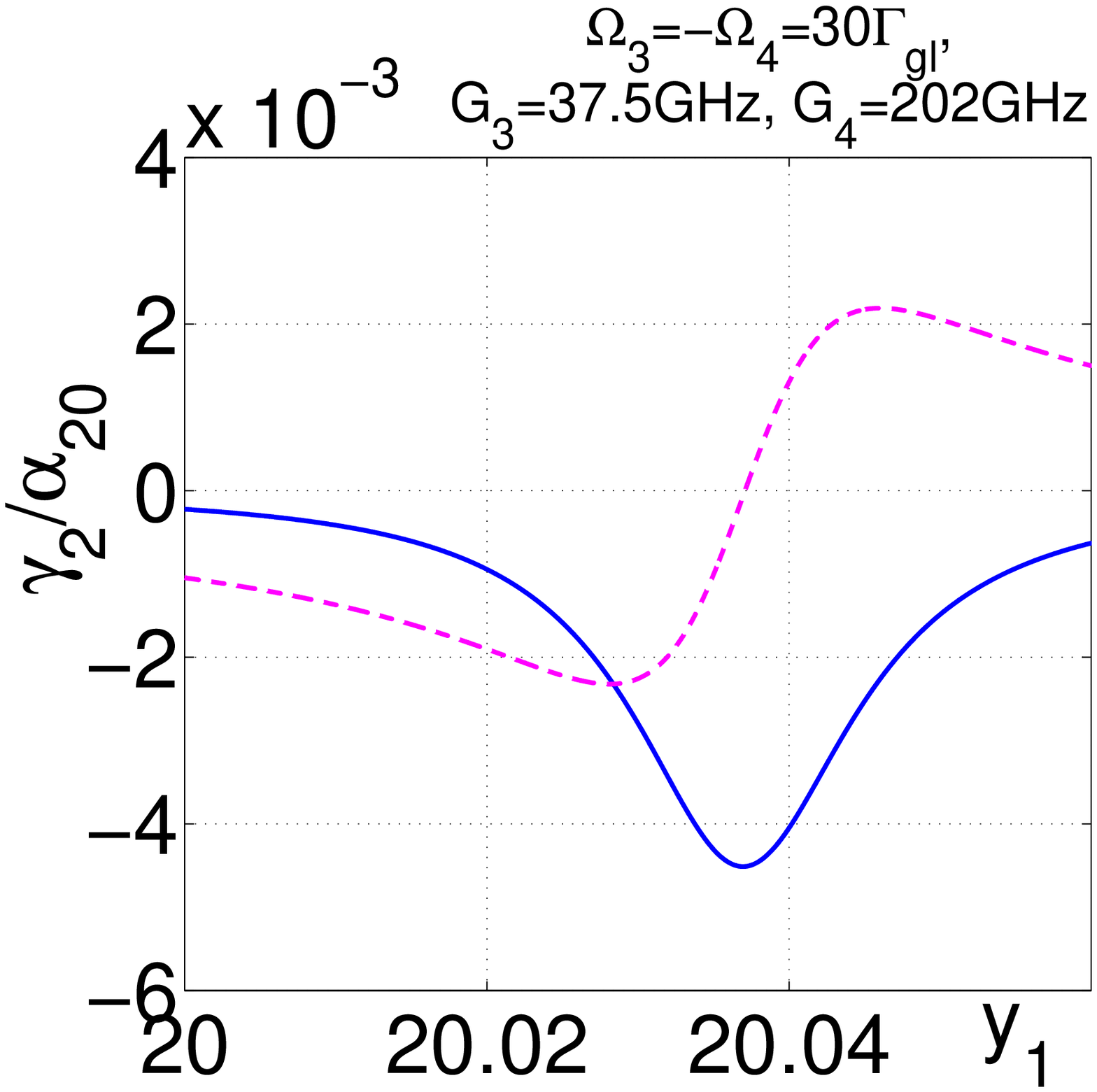 }\\
(c)\hspace{30mm} (d)\\
\includegraphics[width=.4\columnwidth]{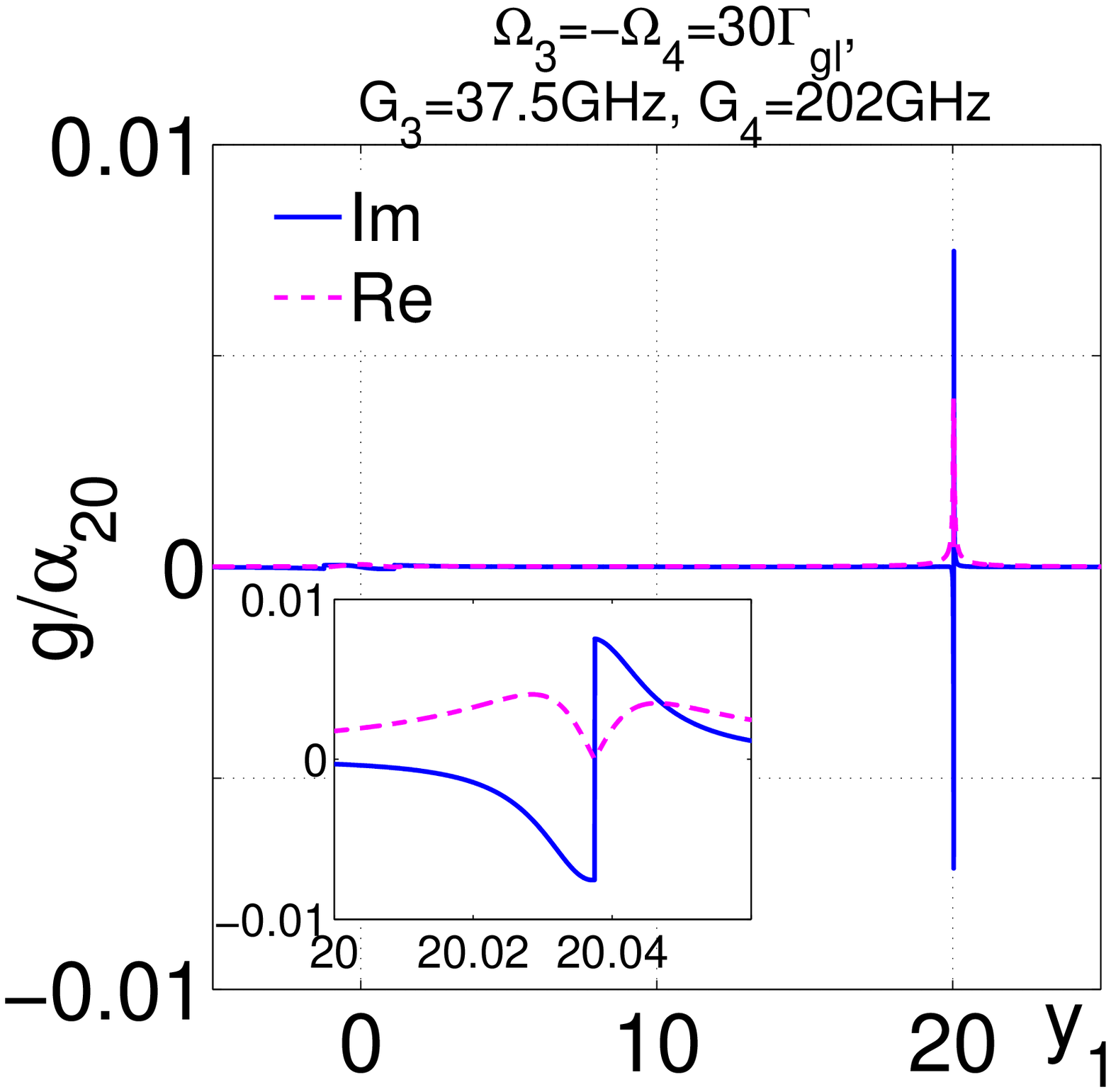}
\includegraphics[width=.4\columnwidth]{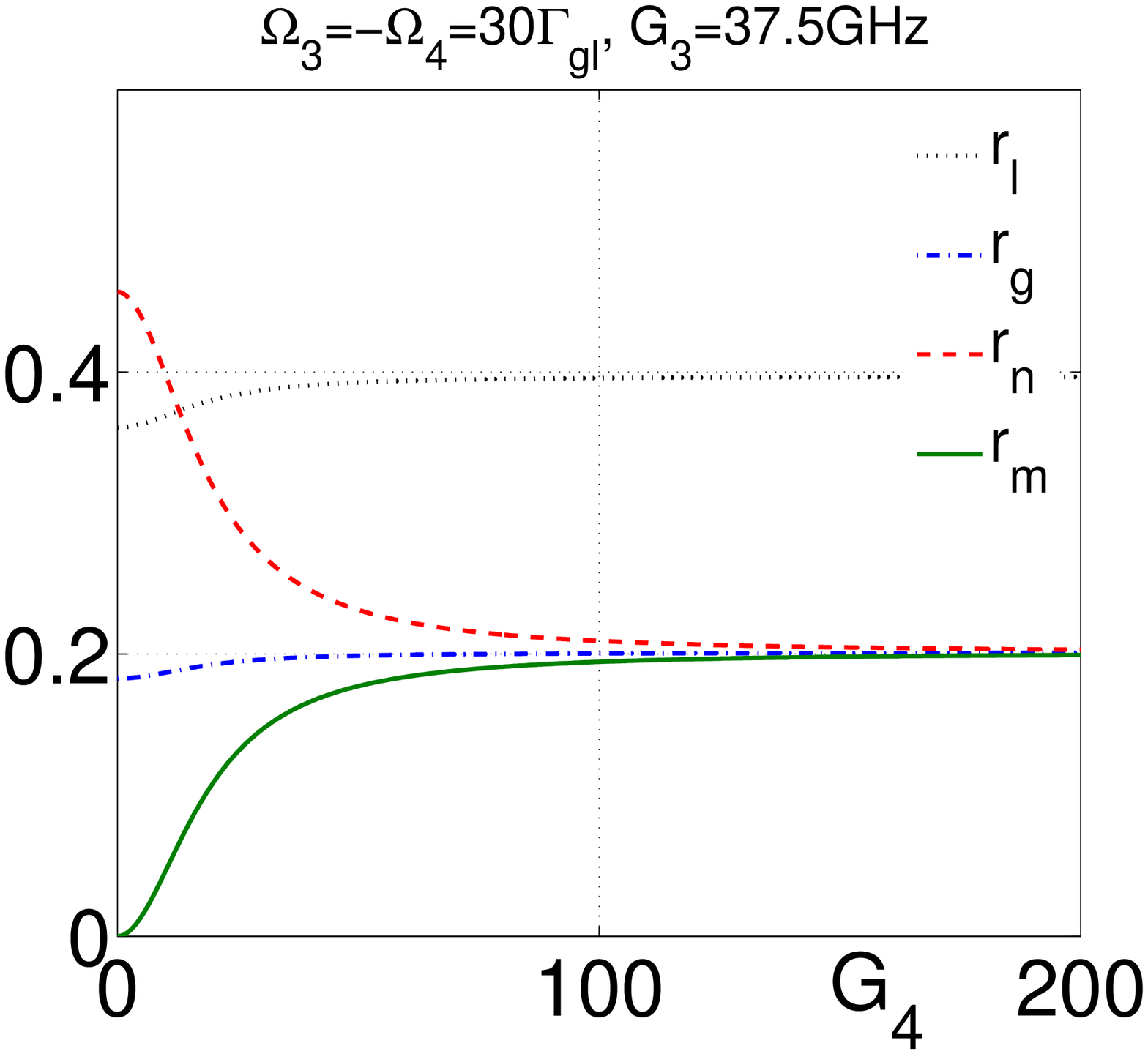}\\
(e)\hspace{30mm} (f)\\
\includegraphics[width=.4\columnwidth]{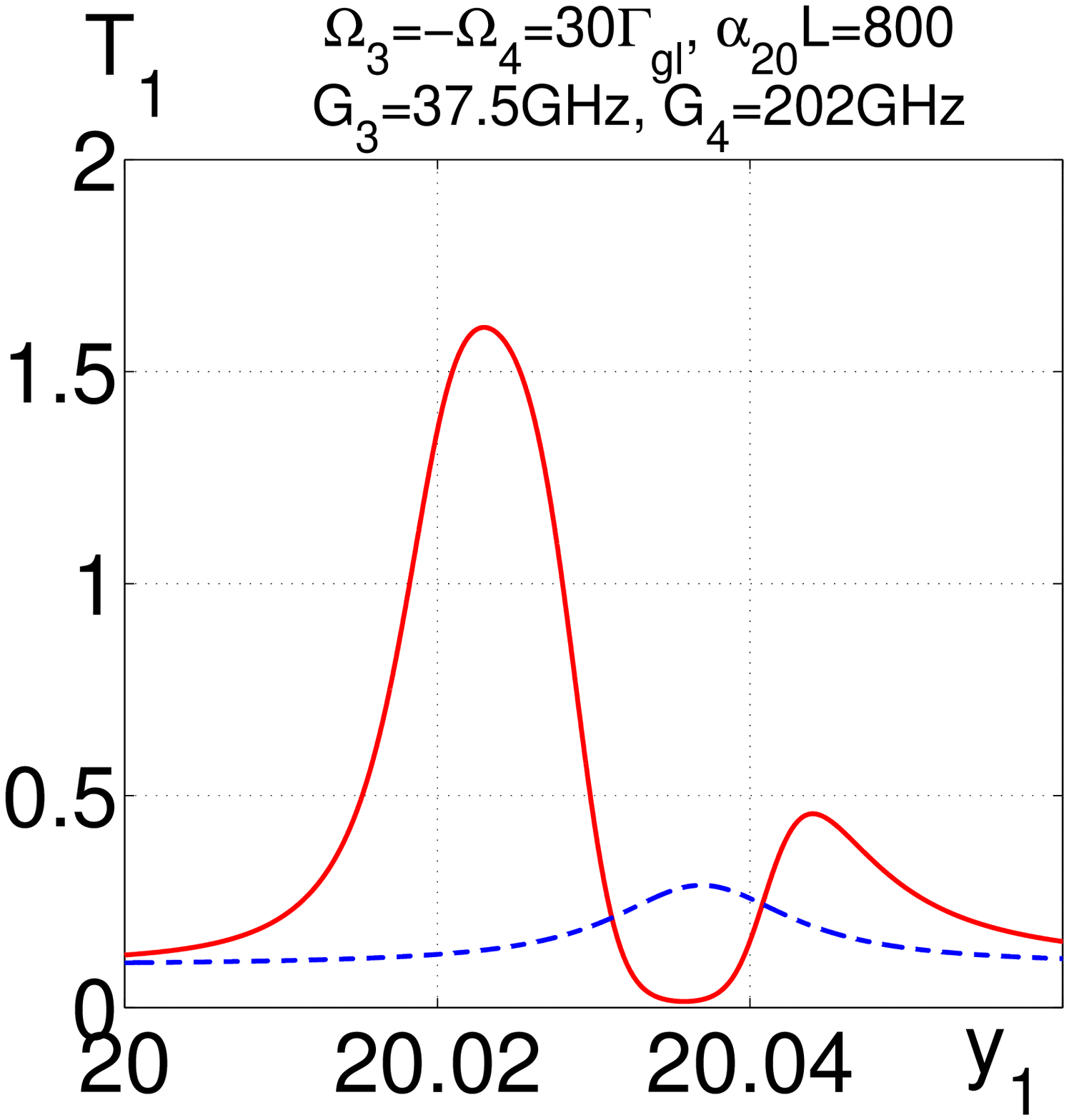}
\includegraphics[width=.4\columnwidth]{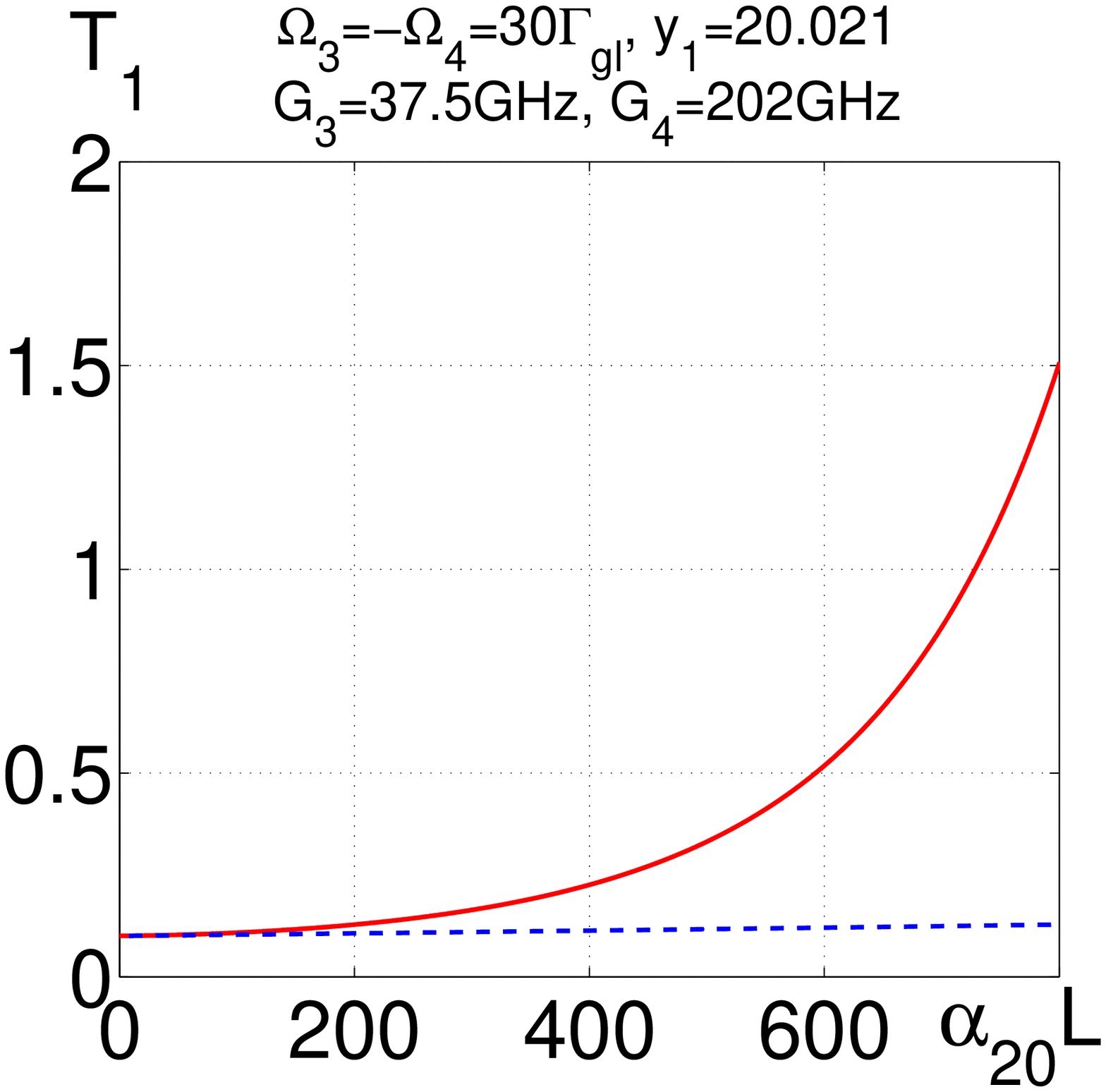}\\
(g)\hspace{30mm} (h)
\end{center}
\caption{Energy-conversion in the scheme with neither population inversion nor two-photon gain possible [$\gamma_{mn}=9\times10^7$ sec$^{-1}$);
all other relaxation parameters are the same as in the previous
case]. $y_1=(\omega_1-\omega_{gn})/\Gamma_{gn}$, $\omega_2= \omega_3+\omega_4-\omega_1$. (a): absorption indices for the signal and the idler; (b):
phase mismatch; (c)-(f): four-wave mixing coupling parameters; (g)
and (h): transmission factor, the dashed line shows transmission at $g=0$.
Coupling Rabi frequencies and
resonance frequency offsets for the control fields are: $G_3$=37.5
GHz, $G_4$=202 GHz, $\Omega_3=-\Omega_4= 30\Gamma_{gl}$.} \label{f7}
\end{figure}
\begin{figure}[!h]
\begin{center}
\includegraphics[width=.38\columnwidth]{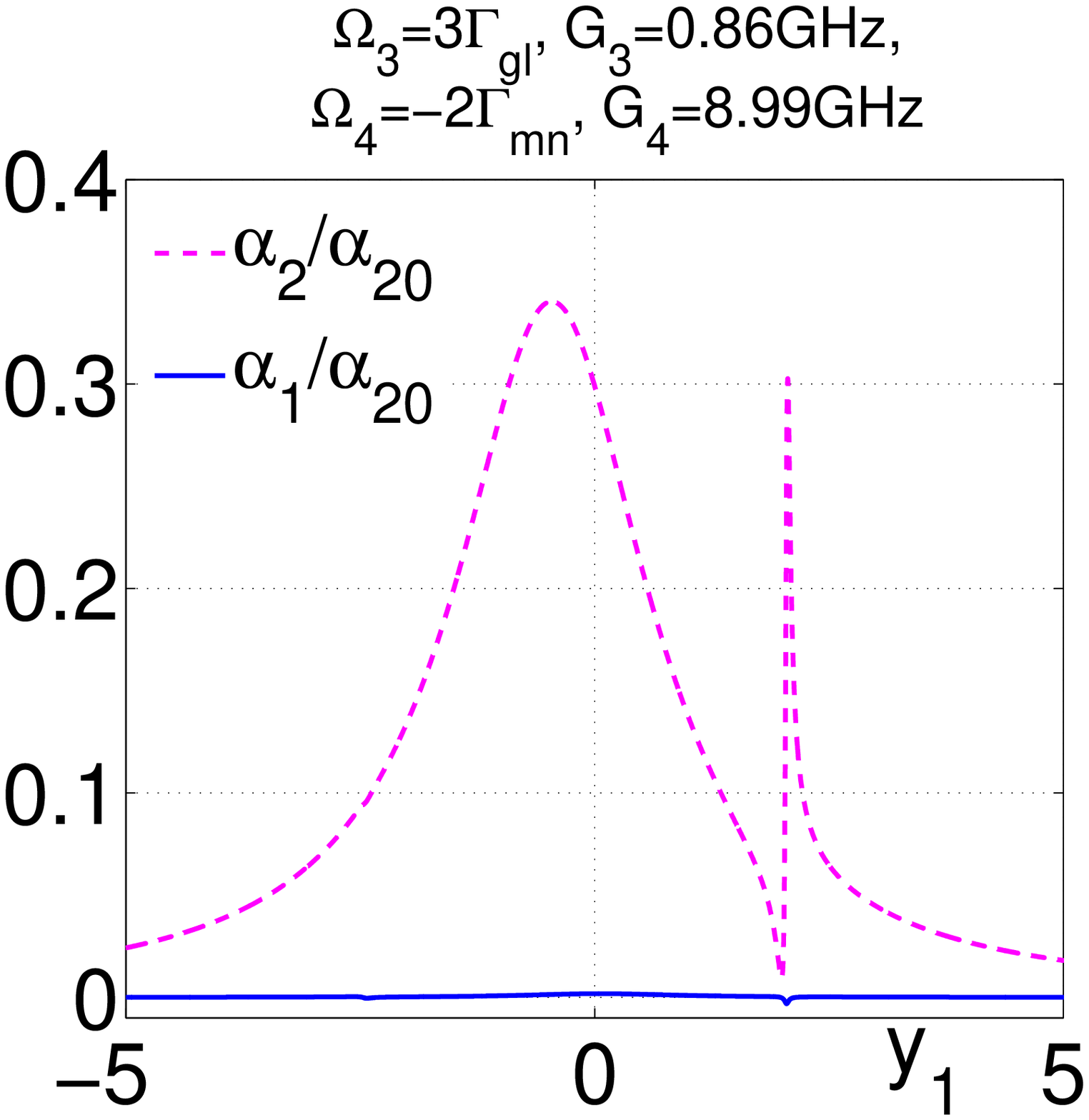}
\includegraphics[width=.38\columnwidth]{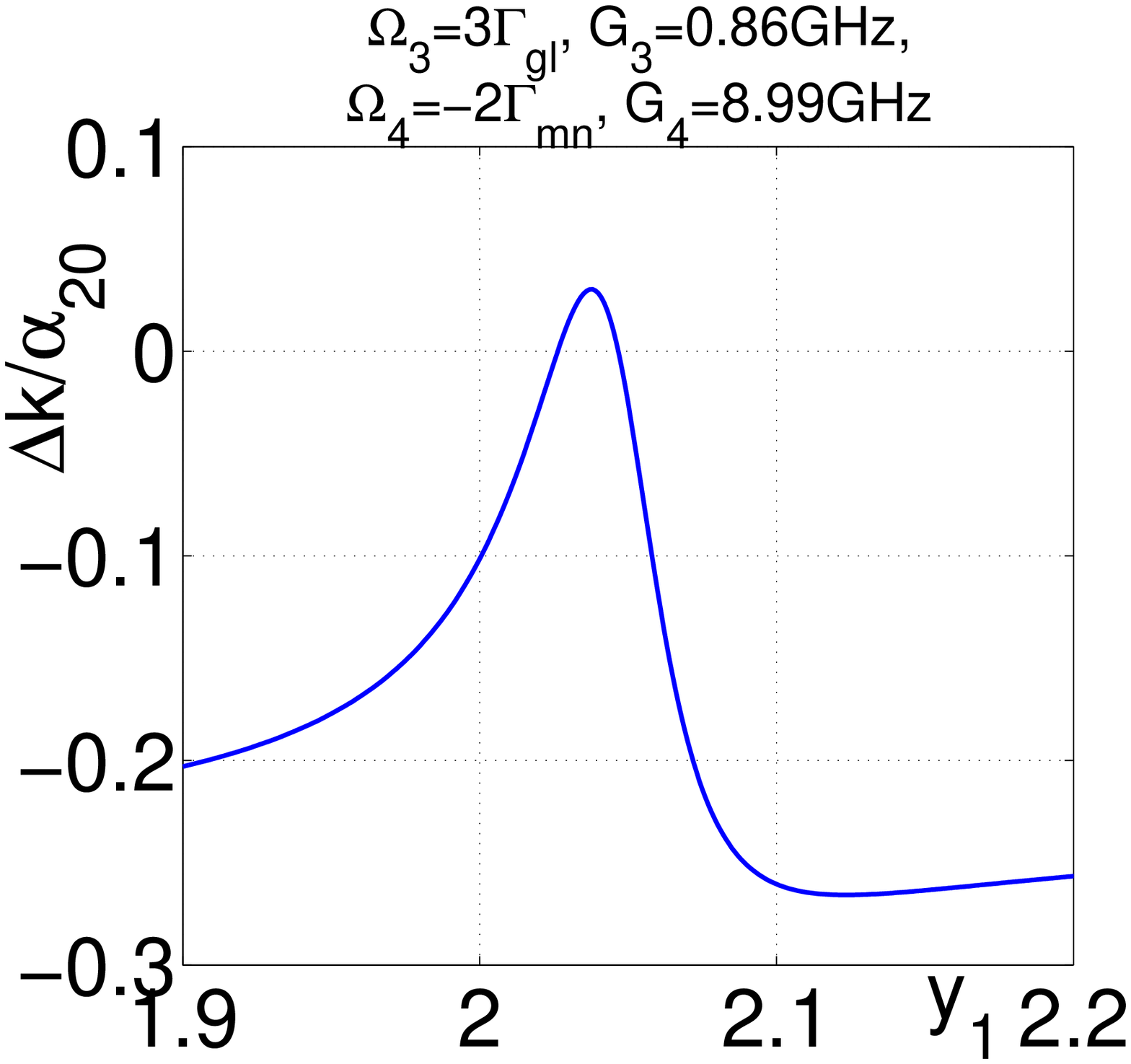}\\
(a)\hspace{30mm} (b)\\
\includegraphics[width=.38\columnwidth]{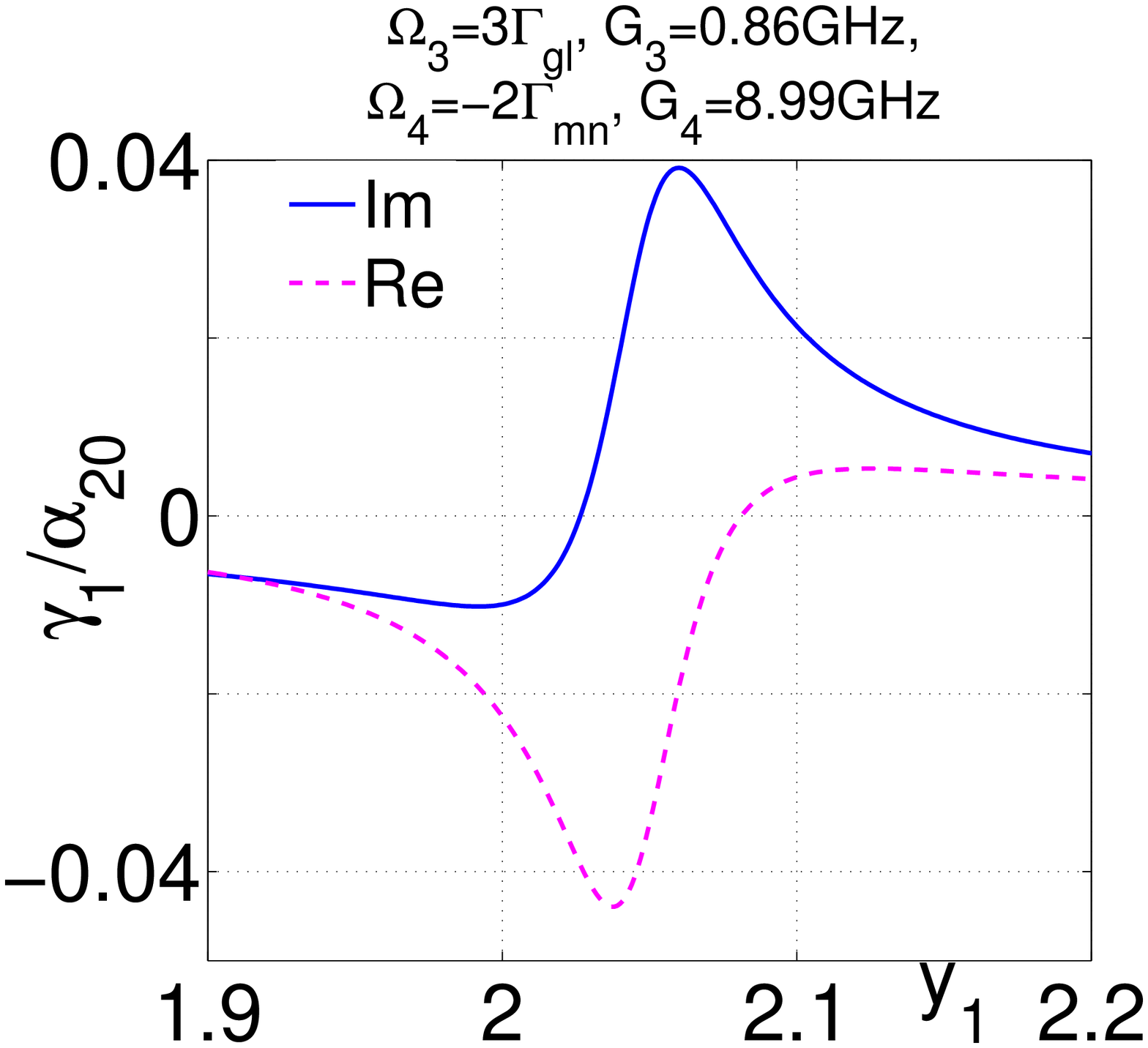}
\includegraphics[width=.38\columnwidth]{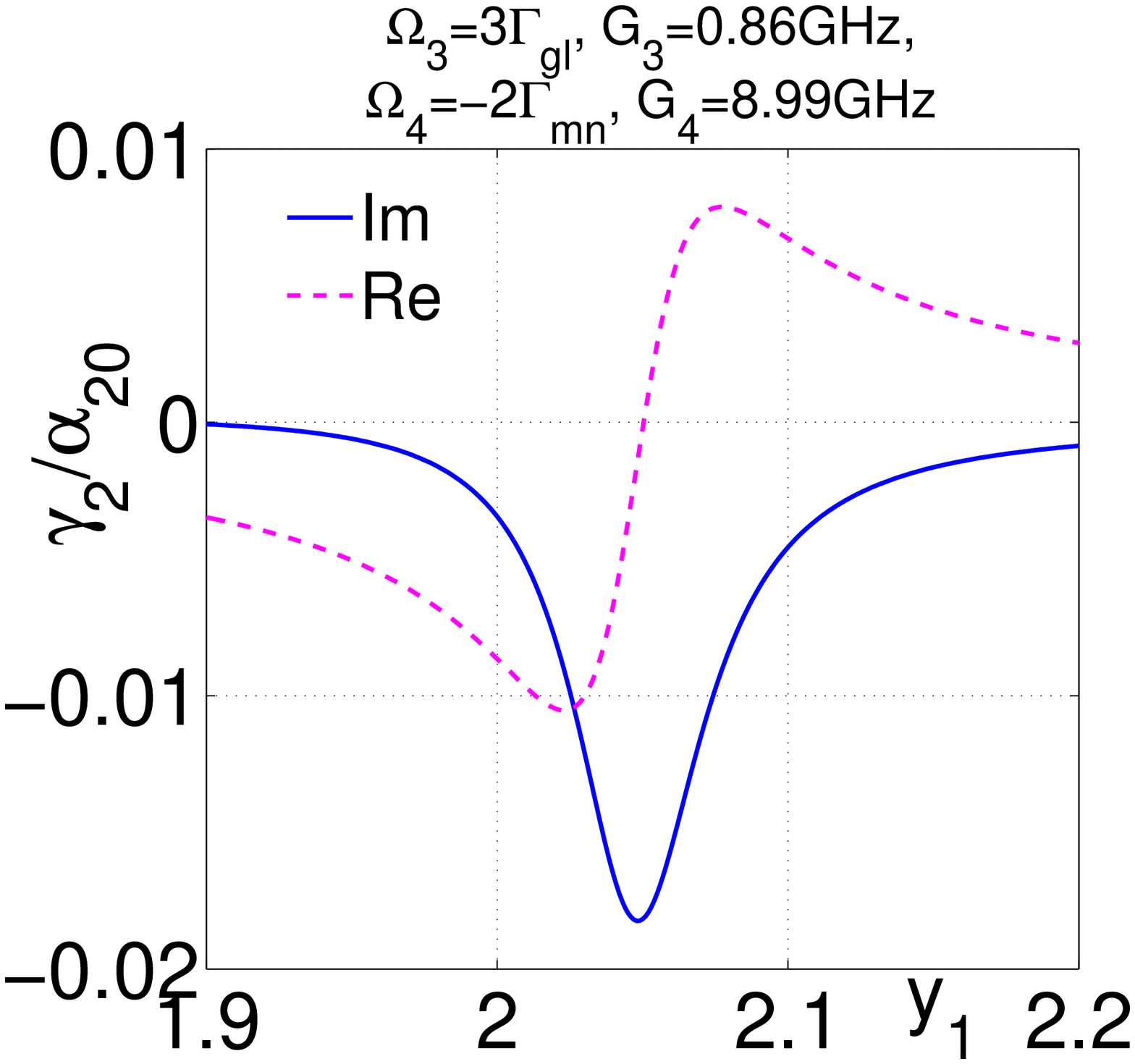}\\
(c)\hspace{30mm} (d)\\
\includegraphics[width=.38\columnwidth]{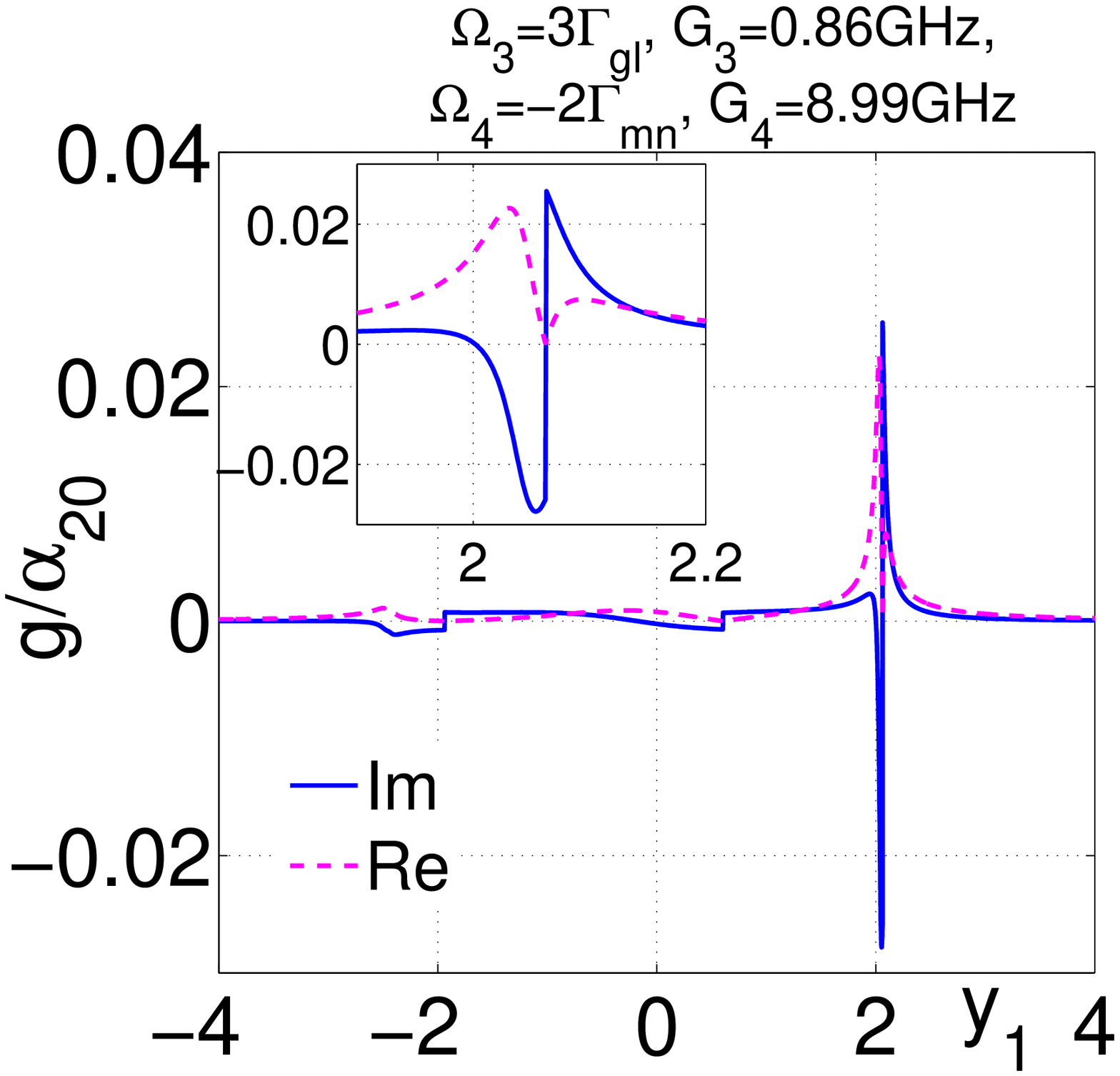}
\includegraphics[width=.38\columnwidth]{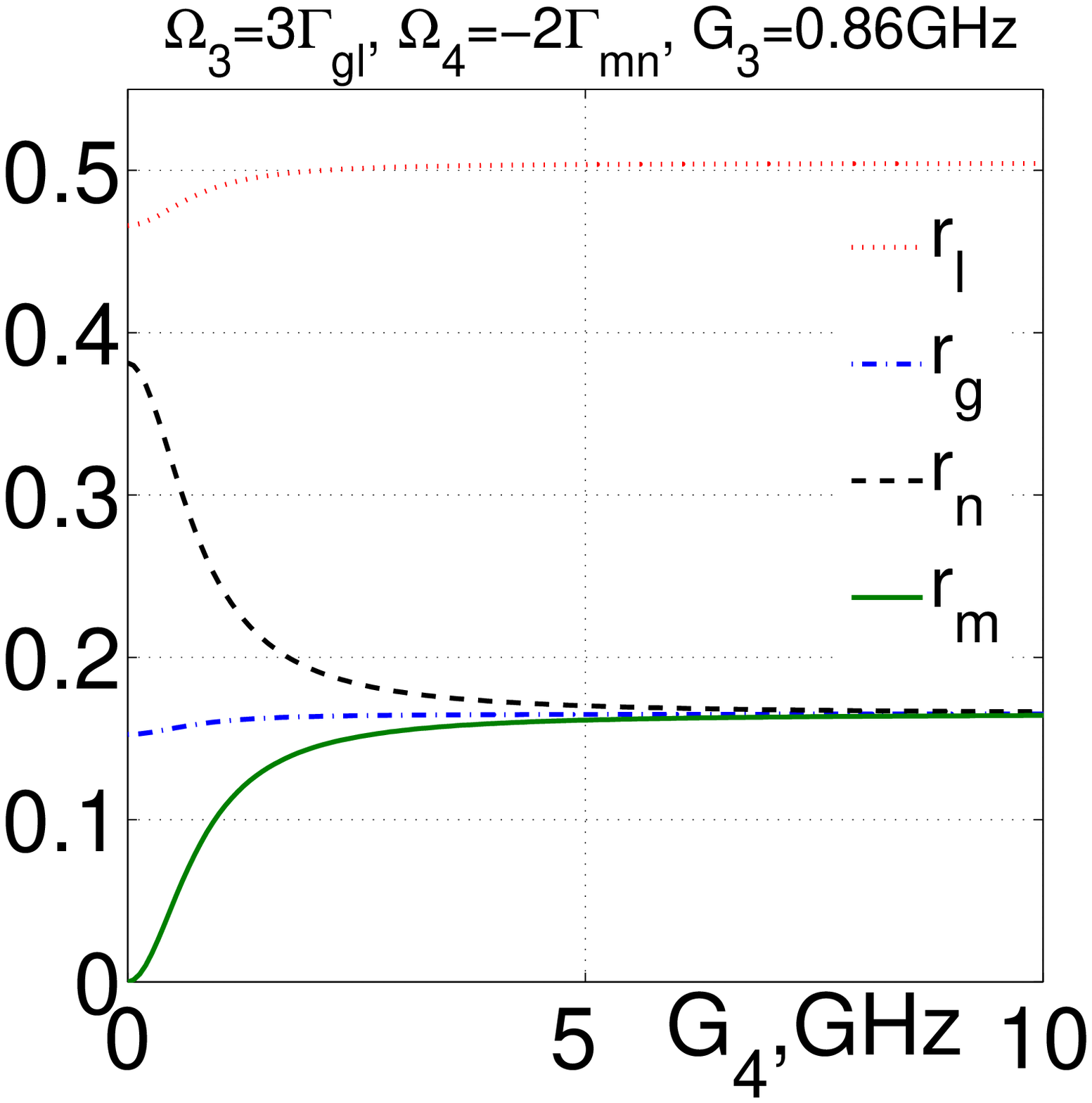}\\
(e)\hspace{30mm} (f)\\
\includegraphics[width=.38\columnwidth]{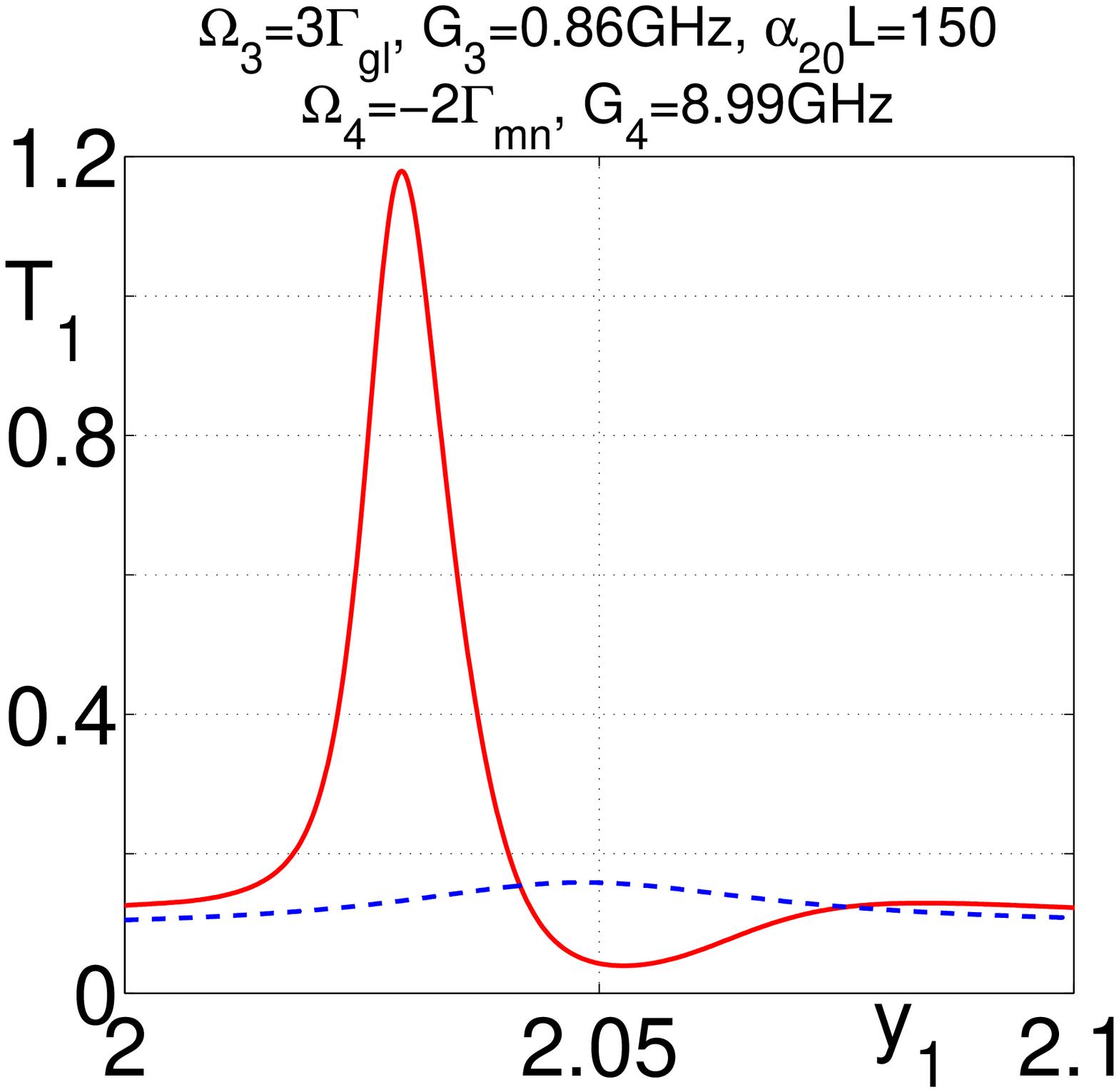}
\includegraphics[width=.38\columnwidth]{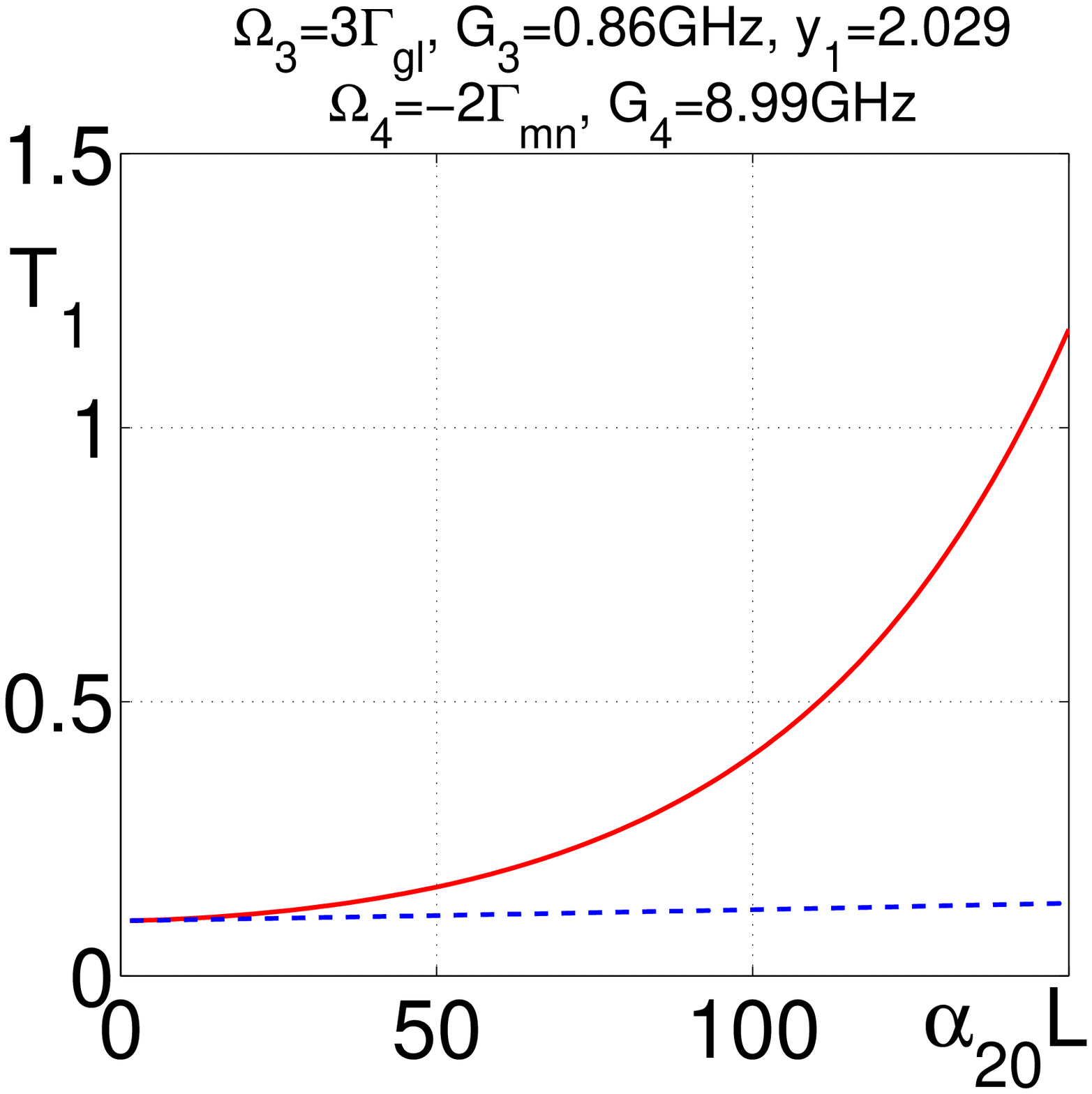}\\
(g)\hspace{30mm} (h)
\end{center}
\caption{Quasi-resonant coupling at lower quantum coherence relaxation
rates and at neither population inversion nor two-photon
gain possible [$\Gamma_{gl}$=1.8, $\Gamma_{mn}$=1.9, $\Gamma_{gn}$=1,
$\Gamma_{ml}$=1.5, $\Gamma_{mg}=5\times10^{-2}$,
$\Gamma_{nl}=5\times10^{-3}$ (in $10^{11}$ sec$^{-1}$); all other
relaxation parameters are the same as in the previous case].   $y_1=(\omega_1-\omega_{gn})/\Gamma_{gn}$, $\omega_2= \omega_3+\omega_4-\omega_1$. (a):
absorption indices for the signal and the idler; (b): phase mismatch;
(c)-(e): four-wave mixing coupling parameters; (f): energy level
populations; (g) and (h): transmission factor, dash line shows
transmission at $g=0$.
Coupling Rabi frequencies and resonance frequency offsets for the
control fields are: $G_3$=0.86 GHz, $\Omega_3=3 \Gamma_{gl}$;
$G_4$=8.99 GHz, $\Omega_4=-2 \Gamma_{mn}$.} \label{f8}
\end{figure}
Figures~\ref{f6}(a)-(d) display the spectral properties of the output signal at $z=0$ for one of the resonances in the vicinity of the signal frequency offset $\omega_1-\omega_{gn}\approx20.05\Gamma_{gn}$ at different optical densities of the slab at $\omega_{ml}$ attributed to the impurity centers. We assume that the absorption of the host material in the slab at $\omega_1$ is fixed at 90\% and it is equal to 88\% at $\omega_2$. The density of the embedded centers and the slab thickness, and hence, the additional resonant optical thickness of the slab contributed by these impurities, may vary as shown in the panels. Actual quasi-resonant absorption/gain indices depend on the intensities and frequency offsets of the control fields, as shown in Fig.\ref{f4}(a). Besides the features imposed by the counter-propagation of the coupled waves, the output magnitudes of the signal at $z=0$ and the idler at $z=L$ and their distributions inside the slab are determined by the interplay of several contributing linear and nonlinear processes. They include the phase mismatch,  absorption of the signal and the idler, and the parametric gain $g$, which are all controlled by the driving fields $E_3$ and $E_4$. The dependence of the overall optimized output signal on the density of the impurities and on the slab thickness  (on the resonant optical thickness of the slab) is depicted in Fig. ~\ref{f6}(e). Such a behavior is determined by the radically different distributions of the idler, which propagates from left to right, and the signal, which propagates from right to left, [Fig. \ref{f6}(f)]. Figures \ref{f6}(b)-(f) indicate the possibility of {mirrorless self-oscillation}. Figure \ref{f7} shows the role of partial spontaneous transitions between the energy levels. Here, $\gamma_{mn}=9\times10^7$ sec$^{-1}$, which makes both population inversion and two-photon gain impossible [Fig. \ref{f7}(a)]. At the indicated Rabi frequencies and frequency offsets for the driving control fields, the energy-level populations are: $r_l \approx 0.4$, $r_g \approx 0.2009$, $r_n\approx0.2031$, $r_m \approx 0.2$. The magnitude of the four-wave mixing coupling parameters appear comparable with those depicted in Fig. \ref{f4}(c)-(h). However, the absence of one- and two-photon amplification that would support energy-conversion processes, like in \cite{APB09} and in Fig. \ref{f4}(a), dramatically decreases the achievable amplification and increases the required optical thickness of the slab [Fig. \ref{f7}(g),(h)]. Figure \ref{f8} shows that, even in such cases, the optimized magnitude of the required control field intensities and the slab optical density can be substantially reduced for centers with lower coherence relaxation rates and quasi-resonant coupling. Here, quantum nonlinear interference effects play an important role \cite{GPRA}.  At the indicated Rabi frequencies and frequency offsets for the driving control fields shown in Fig. \ref{f8}, the energy-level populations are: $r_l \approx 0.504$, $r_g \approx 0.165$, $r_n\approx0.167$, $r_m \approx 0.164$. Like in the previous examples, the losses in the host NIM material are taken to be fixed and equal to $\alpha_{NIM1}L=2.3$ for the signal and $\alpha_{NIM2}L=2.1$ for the idler.

The dependencies presented in Figs. \ref{f4} - \ref{f8} correspond to the
vicinity of the first geometrical resonance, which appears at the lowest
magnitude of $gL$ as shown in Fig.~\ref{f3}. For the above-indicated
characteristics of the involved optical transitions, the magnitude
G=150 GHz corresponds to control field intensities on the order of
of $I\sim$ 10 kW/(0.1mm)$^2$. Assuming a resonance absorption
cross-section $\sigma_{40}~\sim~10^{-16}$~cm$^2$, which is typical
for transitions with oscillator strength of about one, and a
concentration of embedded centers $ N~\sim~10^{19}$ cm$^{-3}$, we
obtain $\alpha_{20}~\sim 10^3$~cm$^{-1}$ and the required slab
thickness in the {microscopic} range $L~\sim (1 - 100) \mu$m. The contribution to the index of refraction by the impurities is estimated as $ \Delta n< 0.5(\lambda/4\pi)\alpha_{40}\sim 10^{-3}$, which essentially does not change the negative refractive index.

\section{Conclusions}
The possibility to produce and to tailor a laser-induced optical
transparency in a negative-index metamaterial slab through
nonlinear-optical coherent energy transfer between ordinary control
wave(s) and a negative-index backward signal is shown and proven by
numerical simulations. Two possible types of nonlinear-optical
couplings are discussed. One is off-resonant three or four-wave mixing
by making use of the nonlinear susceptibilities that are assumed
attributed to the metamaterial nanostructures and are independent of the
intensities and frequencies of the coupled optical fields. The other option is the independent engineering of a resonantly enhanced, four-wave
mixing nonlinearity associated with nonlinear-optical centers
embedded in a negative-index host matrix. In the latter case, the
proposed coupling scheme suggests that the frequency of the negative-index signal should fall in the vicinity of the transition between the excited
levels of the centers, while the idler frequency appears coupled
with the absorptive transition from the ground state. The scheme
under investigation is different from the earlier proposed schemes
and exhibits essentially different features. The extraordinary
properties of the nonlinear-optical propagation processes in both outlined types of  metamaterials are investigated. These properties are in
drastic contrast with their counterparts in ordinary, positive index
materials. The focus of this work is on the possibility of compensating for the strong losses inherent to metal-dielectric negative-index
metamaterials and on producing laser-induced optical transparency and
gain for the negative-index signal. In the case of a frequency- and
intensity-independent nonlinear-optical response of the composite,
the feasibility of producing transparency and amplification through
the entire negative-index frequency domain above a certain control
laser field intensity is investigated. This is shown to be possible by
adjusting the absorption index for the idler to be greater than that
for the negative-index signal. Specific features of the quantum
control attributed to the second scheme are investigated that allow
for transformable optics through  frequency-tunable narrow-band transparency, quantum switching, filtering and amplification of light.  The possibility of compensating the strong losses inherent to NIMs and
realizing a miniature, mirrorless optical parametric generator of
entangled contra-propagating backward and ordinary waves is also
shown and supported by numerical simulations. The extraordinary
features predicted in this work stem from the backwardness of
electromagnetic waves, which is a feature inherent to this type of
metamaterial.

\section*{Acknowledgment}
This work was supported by the U.~S. Army Research Laboratory and by
the U. S. Army Research Office under grants number W911NF-0710261 and
50342-PH-MUR and by the Siberian Division of the Russian Academy of
Sciences under Integration Project No 5.

\end{document}